\documentclass[twocolumn, pra, superscriptaddress]{revtex4}

\usepackage{amsmath}
\usepackage{amsfonts}
\usepackage{bbm}
\usepackage{graphicx}
\usepackage[colorlinks]{hyperref}
\usepackage{mathtools}
\usepackage{mathrsfs}
\usepackage{physics}
\usepackage{datetime2}

\usepackage{nicefrac}
\usepackage[capitalise]{cleveref}
\usepackage[dvipsnames]{xcolor}

\usepackage[normalem]{ulem}

\def \addCQuIC {Center for Quantum Information and Control, University
  of New Mexico, Albuquerque, NM, USA}

\def \addCQuIC {Center for Quantum Information and Control, University of New Mexico, Albuquerque, NM, USA}
\def \addSandia {Sandia National Laboratories, Albuquerque, NM, USA}
\def \addPandAUNM {Department of Physics and Astronomy, University of New Mexico, Albuquerque, NM, USA}
\def \addPandAUO {Department of Physics and Astronomy, University of Oklahoma, Norman, OK, USA}
\def \addLANL {Los Alamos National Laboratory, Los Alamos, NM, USA}

\begin{document}

\title{Robust M{\o}lmer-S{\o}rensen gate for neutral atoms using rapid adiabatic Rydberg dressing}

\author{Anupam Mitra}
\email{anupam@unm.edu}
\affiliation{\addCQuIC} \affiliation{\addPandAUNM}

\author{Michael J. Martin}
\affiliation{\addLANL} \affiliation{\addSandia}

\author{Grant W. Biedermann}
\affiliation{\addCQuIC} \affiliation{\addSandia} \affiliation{\addPandAUO}

\author{Alberto M. Marino}
\affiliation{\addPandAUO} \affiliation{\addSandia}

\author{Pablo M. Poggi}
\affiliation{\addCQuIC} \affiliation{\addPandAUNM}

\author{Ivan H. Deutsch}
\email{ideutsch@unm.edu}
\affiliation{\addCQuIC} \affiliation{\addPandAUNM}

\date{2020-02-14}

\begin{abstract}
The Rydberg blockade mechanism is now routinely considered for entangling qubits encoded in clock states of neutral atoms. Challenges towards implementing entangling gates with high fidelity include errors due to thermal motion of atoms, laser amplitude inhomogeneities, and imperfect Rydberg blockade. We show that adiabatic rapid passage by Rydberg dressing  provides a mechanism for implementing two-qubit entangling gates by accumulating phases that are robust to these imperfections. We find that the typical error in implementing a two-qubit gate, such as the controlled phase gate, is dominated by errors in the single atom light shift, and that this can be easily corrected using adiabatic dressing interleaved with a simple spin echo sequence. This results in a two-qubit M{\o}lmer-S{\o}rensen gate. A gate fidelity $\sim 0.995$ is achievable with modest experimental parameters and a path to higher fidelities is possible for Rydberg states in atoms with a stronger blockade, longer lifetimes, and larger Rabi frequencies.
\end{abstract}

\maketitle


Arrays of trapped neutral atoms interacting via the electric dipole-dipole interaction (EDDI) have emerged as a potential scalable platform for quantum computing~\cite{brennen1999quantum, brennen2000entangling, weiss2017quantum}. Near term applications include simulation of Ising models \cite{labuhn2016tunable, de2018accurate} and optimization~\cite{keating2013adiabatic, zhou2018quantum, pichler2018computational}. In the longer term, this system is a promising platform for universal fault-tolerant quantum computing given long-lived qubits at the heart of ultraprecise atomic clocks~\cite{ludlow2015optical} and flexible trapping geometries \cite{endres2016atom, barredo2016atom, barredo2018synthetic, kumar2018sorting}. High fidelity, on demand one-qubit gates have been demonstrated~\cite{wang2015coherent, wang2016single} but the implementation of two-qubit gates with the fidelities required for fault tolerance remains a critical challenge.

 Entangling gates based on the EDDI Rydberg-blockade mechanism~\cite{jaksch1999entanglement, lukin2001dipole, saffman2010quantum, saffman2016quantum} were first demonstrated in seminal experiments~\cite{wilk2010entanglement, isenhower2010demonstration}. In recent developments, high fidelity entangling interactions of qubits encoded in ground and Rydberg states ~\cite{levine2018high} have been applied  in a variety of applications ~\cite{bernien2017probing, keesling2019quantum, omran2019generation}, and the controlled-Z (CZ) gate on clock-state qubits have been demonstrated with a fidelity $\sim 0.97$ in 1D~\cite{levine2019parallel} and $\sim 0.89$ in 2D~\cite{graham2019rydberg} arrays.
 
To achieve higher fidelity two-qubit entangling gates, we consider dressing clock states with Rydberg states via adiabatic rapid passage, a powerful tool for robust control~\cite{kral2007colloquium}. Rydberg dressing has been studied for application in simulation~\cite{pohl2003plasma, johnson2010interactions, zeiher2016many} and metrology ~\cite{gil2014spin, norcia2019seconds,pupillo2010strongly, dauphin2012rydberg, keating2013adiabatic, dauphin2012rydberg, keating2016arbitrary}. We have employed strong Rydberg dressing to create two-qubit entangled states~\cite{jau2016entangling}, measured the light-shifts of the adiabatically dressed entangled states~\cite{lee2017demonstration}, and showed how adiabatic dressing can be employed to implement a CZ gate with the potential for Doppler-free excitation~\cite{keating2015robust}. Adiabatic passage has also been studied in a variety of protocols as a mechanism for achieving entangling gates with the Rydberg blockade mechanism ~\cite{moller2008quantum, rao2014robust, beterov2016two, wu2017rydberg, beterov2018adiabatic, saffman2019symmetric}. 

In the current work we extend our analysis and show how adiabatic passage to implement Rydberg dressing facilitates a method for realizing a M{\o}lmer-S{\o}rensen (MS) gate ~\cite{molmer1999multiparticle, sorensen1999quantum}, with intrinsic robustness to a wide variety of imperfections. These include inhomogeneities in intensity and in detuning, such as those arising from Doppler shifts at finite temperature and Stark shifts from stray electric fields. As we will show, the dominant effect of such inhomogeneities is the errors incurred by the single atom light shifts, which can be removed using a simple spin echo. Moreover, we can achieve adiabatic rapid passage such that the integrated time spent in the Rydberg state is on the same order as that for the standard pulsed protocol of Jaksch \emph{et al.}~\cite{jaksch2000fast} with equivalent Rabi frequencies, thereby maintaining a similar budget in the error due to finite Rydberg radiative lifetime. Gate fidelities $\sim 0.995$ are compatible with the typical inhomogeneities in current experiments and Rydberg state lifetimes $t \sim 100 \mu s$. Longer-lived Rydberg states can faciliate better control of adiabatic passage to push fidelities even higher.


\begin{figure}
    \centering
    \includegraphics[width=0.48\textwidth]{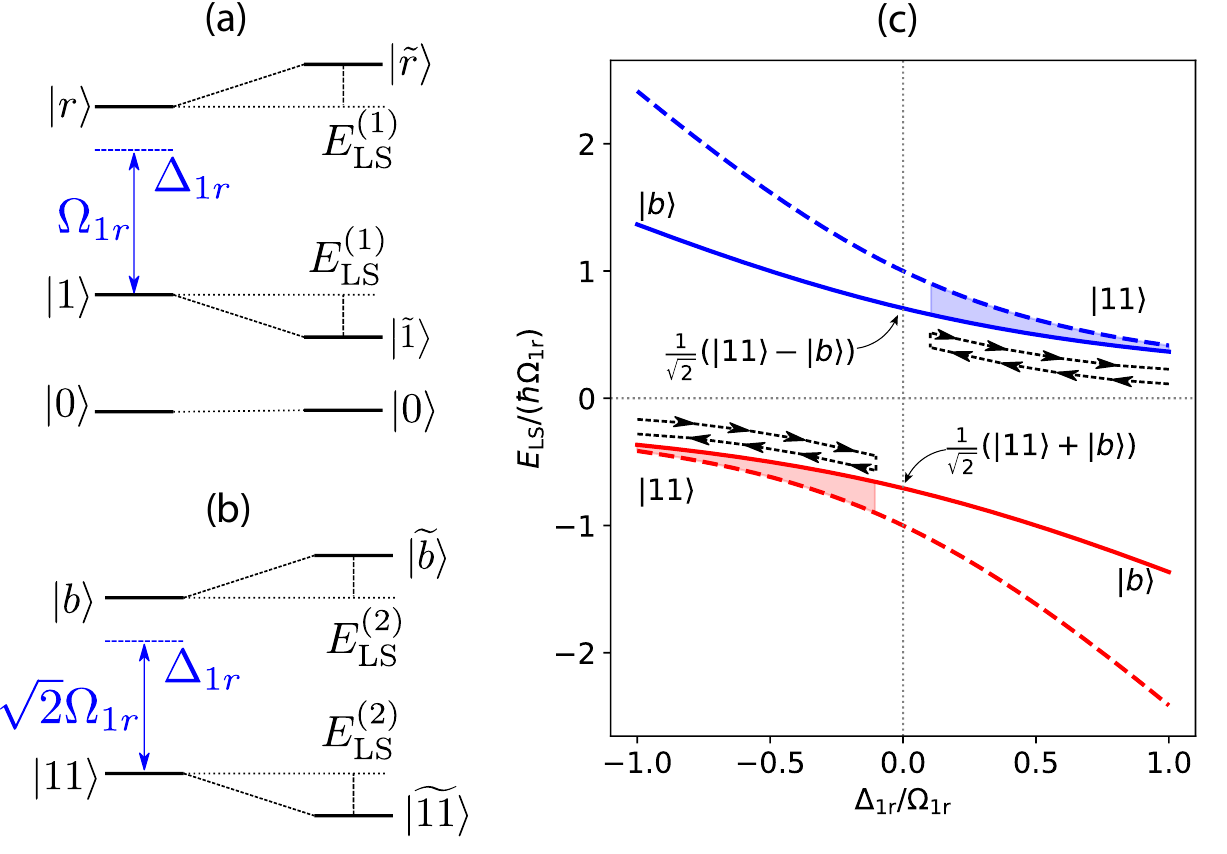}
    \caption{(a) Qubit encoded into atomic clock states with the upper clock state, $|{1}\rangle$, coupled to a Rydberg state $|{r}\rangle$ with a Rabi frequency $\Omega_{1r}$ and detuning $\Delta_{1r}$.
    (b) The two-atom state $|{11}\rangle$ is coupled to the entangled bright state $|{b}\rangle = (\ket{1r}+\mathrm{e}^{\mathrm{i} \varphi}\ket{r1})/\sqrt{2}$ with Rabi frequency $\sqrt{2} \Omega_{1r}$.
    (c) Light shift of the state $|{11}\rangle$ as a function detuning. The dashed lines show the light shift in the absence of EDDI (top: starting from blue detuning, bottom: starting from red detuning), in an adiabatic passage to the doubly excited Rydberg state $|{rr}\rangle$. The solid lines show the light shift in the presence of EDDI under the perfect blockade approximation (top: starting from blue detuning, bottom: starting from red detuning). The shaded region shows the value of $\kappa$ (\cref{eq:kappa}). The dotted lines with arrows show examples of adiabatic passages used in our ramps (\cref{fig:AdiabaticRamp}) to obtain entanglement in the ground state.
    }
    \label{fig:RydbergDressing}
\end{figure}

For generality, we consider an atom with two long-lived clock states to serve as the qubit states $\ket{0}$ and $\ket{1}$. These could be the hyperfine clock states of alkali atoms separated by a microwave frequency (e.g., Cs or Rb) ~\cite{saffman2016quantum}, or the optical clock states of alkaline earth-like atoms (e.g., Sr or Yb) ~\cite{norcia2019seconds, covey20192000, saskin2019narrow}. The clock state $\ket{1}$ is optically coupled to a high-lying Rydberg state $\ket{r}$. The fundamental Hamiltonian governing the Rydberg gate between two atoms is $\hat{H}_{\text{2atom}} = \hat{H}_\alpha + \hat{H}_\beta + V_{\text{DD}}|{rr}\rangle\langle{rr}|$ where $V_{\text{DD}}$ is the electric dipole interaction. $\hat{H}_\alpha$ ($\hat{H}_\beta$) is the Hamiltonian for the atom $\alpha$ ($\beta$) coupled to the Rydberg laser, 
\begin{align}
    \label{eq:OneAtomRydbergHamiltonian}
    \hat{H}_{\alpha} = \tfrac{1}{2m}\hat{p}^2_\alpha-\Delta_{1r} |{r}\rangle_\alpha \langle{r}|_\alpha
    + \tfrac{1}{2}\Omega_{1r}\left( \mathrm{e}^{\mathrm{i} k_r \hat{z}_\alpha} |{r}\rangle_\alpha \langle{1}|_\alpha
    + \text{h.c.} \right),
\end{align}
where $\Omega_{1r}$ and $\Delta_{1r}$ are the Rydberg laser Rabi frequency and detuning respectively (here and throughout we set $\hbar=1$), $\hat{p}_\alpha$ is the atomic momentum operator and $\hat{z}_\alpha$ is the atomic position operator in the direction of the Rydberg laser. Ideally, the atoms are illuminated uniformly and see the same Rydberg laser intensity and detuning. Under those conditions it is natural to consider the basis of ``bright" $\ket{b}=\qty(\mathrm{e}^{\mathrm{i} k_{1r} z_\beta} \ket{1r} + \mathrm{e}^{\mathrm{i}k_{1r} z_\alpha} \ket{r1}) / \sqrt{2}$ and ``dark" $\ket{d}=\qty(\mathrm{e}^{\mathrm{i}k_r z_\beta} \ket{1r} - \mathrm{e}^{\mathrm{i}k_r z_\alpha} \ket{r1})/ \sqrt{2}$ states (here and throughout we use the abbreviated notation, $\ket{xy} = \ket{x}_\alpha \ket{y}_\beta$). We consider Rydberg interactions for atoms released from a trap, as is typically done in experiments and treat the motion as that of a free particle, and the strong blockade regime $V_{\text{DD}}\gg \Omega_{1r}$ and neglect to zeroth order any population in the doubly excited Rydberg state $|{rr}\rangle$; corrections will be considered below. In the case of ``frozen'' atoms with zero momentum, the two-atom Hamiltonian takes the form \cite{johnson2010interactions, keating2015robust}
\begin{align}
  \label{eq:TwoAtomRydbergHamiltonian}
    \hat{H}_{\text{2atom}}=
    -\Delta_{1r} (\ketbra{b}{b}+\ketbra{d}{d}
    + \frac{\sqrt{2} \Omega_{1r}}{2}\left( \ketbra{b}{11} + \text{h.c.} \right).
\end{align}
A $\pi$-rotation on the $\ket{11}\rightarrow\ket{b}$ transition yields an entangled state, recently achieved with fidelity $97\%$~\cite{levine2018high}. When there is thermal motion, the relative phase $\mathrm{e}^{\mathrm{i} k_{1r} (z_\beta-z_\alpha)}$ will vary, leading to coupling between bright and dark states, which limits the transfer the entanglement from the bright state to the long-lived ground state qubits \cite{wilk2010entanglement, graham2019rydberg}.

The dressed states of the zero-momentum two-atom Hamiltonian are ~\cite{keating2015robust}
\begin{align}
    \ket*{\widetilde{11}}
    = \cos{\frac{\theta_2}{2}} \ket{11} + \sin{\frac{\theta_2}{2}} \ket{b} \\
     \ket*{\widetilde{b}}
    = \cos{\frac{\theta_2}{2}} \ket{b} - \sin{\frac{\theta_2}{2}} \ket{11},
\end{align}
where $\tan{\theta_2} = -\sqrt{2}\Omega_{1r}/\Delta_{1r}$. In the dressed states, some character of the entangled bright state $\ket{b}$ is admixed with the  ground state $\ket{11}$. The two-atom light-shift of the ground state, mediated by the Rydberg blockade, is a shift in the energy eigenvalues of the dressed states with respect to the bare states, which under the perfect blockade approximation is $E_{\text{LS}}^{(2)} = \frac{1}{2}\qty(-\Delta_{1r} \pm \sqrt{2 \Omega_{1r}^2+\Delta_{1r}^2})$ \cite{johnson2010interactions, keating2015robust, jau2016entangling, lee2017demonstration}. In the absence of the EDDI the light shift of this state is equal to twice the single atom shift, $2 E^{(1)}_{\mathrm{LS}}$.  There difference between the interacting and noninteracting shifts is the \textit{entangling energy} $\kappa$ ~\cite{keating2015robust, jau2016entangling, lee2017demonstration},
\begin{align}
  \label{eq:kappa}
    \kappa
    &= E_{\mathrm{LS}}^{(2)}
    - 2 E_{\mathrm{LS}}^{(1)} \nonumber
    \\
    &= \frac{1}{2} \qty(\Delta_{1r}
    \pm 
    \qty(\sqrt{2 \Omega_{1r}^2 + \Delta_{1r}^2}
    - 2 \sqrt{\Omega_{1r}^2 + \Delta_{1r}^2})).
\end{align}
On resonance $\kappa \approx \pm 0.29 \Omega_{1r}$, where $\Omega_{1r}/2\pi$ can be as large as a few MHz. For weak dressing, $|\Delta_{1r}|\gg \Omega_{1r}$,  $\kappa \approx - \Omega_{1r}^4/8\Delta^3_{1r}$, which will generally be smaller than the rate of photon scattering, which scales as $1/\Delta^2_{1r}$. Thus weak dressing will not yield high fidelity entangling gates in our protocol.

The dressed energy levels provide an adiabatic passage from the one-atom ground state $\ket{1}$ to the one-atom Rydberg state $\ket{r}$ and from the two-atom ground state $\ket{11}$ to the two-atom entangled bright state $\ket{b}$, as shown in \cref{fig:RydbergDressing}(c). Assuming adiabatic evolution, we consider sweeping the detuning from $\ket{11}$ toward $\ket{b}$ and then back to $\ket{11}$, yielding an entangling phase given by $\vartheta_2= \int \kappa \dd t$. While $\kappa$ grows monotonically as we pass adiabatically from $\ket{11}$ to $\ket{b}$, the simultaneous restrictions of maintaining adiabaticity and limiting the phase $\vartheta_2$ to the target value puts a constraint on the value of the final detuning. Operationally this final detuning is near resonance in our protocol, yielding the minimum gate time such that we simultaneously remain adiabatic but act fast compared to the decoherence rates.

To understand the general class of gates enabled by the phases accumulated in adiabatic evolution and their sensitivity to errors, we consider the Hamiltonian in the dressed qubit (DQ) ground subspace $\qty{\ket{00}, \ket*{0\tilde{1}}, \ket*{\tilde{1}0}, \ket*{\widetilde{11}}}$,  where $\ket*{\tilde{1}}$ is the one-atom dressed ground state that is a superposition of $\ket{1}$ and $\ket{r}$ with dressing angle given by $\tan \theta_1 = -\Omega_{1r}/\Delta_{1r}$. Let $\hat{\sigma}_z = \ketbra*{\tilde{1}}{\tilde{1}} - \ketbra{0}{0}$ be the adiabatic Pauli operator on one atom and  $\hat{S}_z = \mathbbm{1} \otimes \hat{\sigma}_z/2 + \hat{\sigma}_z/2 \otimes \mathbbm{1}$ be the collective angular momentum operator. In the dressed atomic basis, the Hamiltonian in the ground subspace can be written as
\begin{align}
    \label{eq:RydbergDressedGroundHammy}
    \hat{H}_{\mathrm{DQ}} =
    - \qty(E_\mathrm{LS}^{(1)} + \frac{\kappa}{2}) \hat{S}_z
    + \kappa \frac{\hat{S}_z^2}{2}.
\end{align}
This Hamiltonian generates symmetric, one axis, two qubit unitary transformations. The $\hat{S}_z$ term generates $\mathbb{SU}(2)$ rotations on the collective spin, while the $\hat{S}_z^2$ term ``twists" the collective spin and also generates two-qubit entanglement. The quantization axis can be changed to any axis $\mu$ using additional global $\mathbb{SU}(2)$ rotations.

Consider, thus, the unitary transformation of the dressed qubits generated by adiabatic evolution with this Hamiltonian,
\begin{align}
    \label{eq:OneAxisTwistRotateUnitary}
    \hat{U}_{\kappa} =
    \exp\qty(-\mathrm{i} \vartheta_1 \hat{S}_{\mu}
    - \mathrm{i} \vartheta_2 \frac{\hat{S}_{\mu}^2}{2}),
\end{align}
where $\vartheta_1 = -\int \left(E_\mathrm{LS}^{(1)} + \frac{\kappa}{2}\right) \dd t $ is the rotation angle generated by the linear term, and $\vartheta_2 = \int \kappa \dd t$ is the twist angle generated by the quadratic term in the Hamiltonian. When the twist angle $\vartheta_2 = \pm \pi$, these gates are perfect entanglers ~\cite{zhang2003geometric}, meaning that the gates can take a product state to a maximally entangled state. Examples of perfect entanglers of this kind are the CZ gate ($\mu = z, \vartheta_1 = \mp \pi/2, \vartheta_2 = \pm \pi$) and the MS gate ($\mu = x, \vartheta_1 = 0, \vartheta_2 = \pi$). A CZ gate is achieved by removing the phases accumulated due to the independent one atom light shifts $E_{\mathrm{LS}}^{(1)}$ \cite{keating2015robust}. In contrast, the MS gate is achieved by removing \textit{all} single qubit phases contributing to $\vartheta_1$. While theoretically this difference is trivial, the dominant source of gate infidelity is errors in $\vartheta_1$, making the MS gate more robust than the CZ gate, as we will see below.

\begin{figure}
    \centering
    \includegraphics[width=0.48\textwidth]{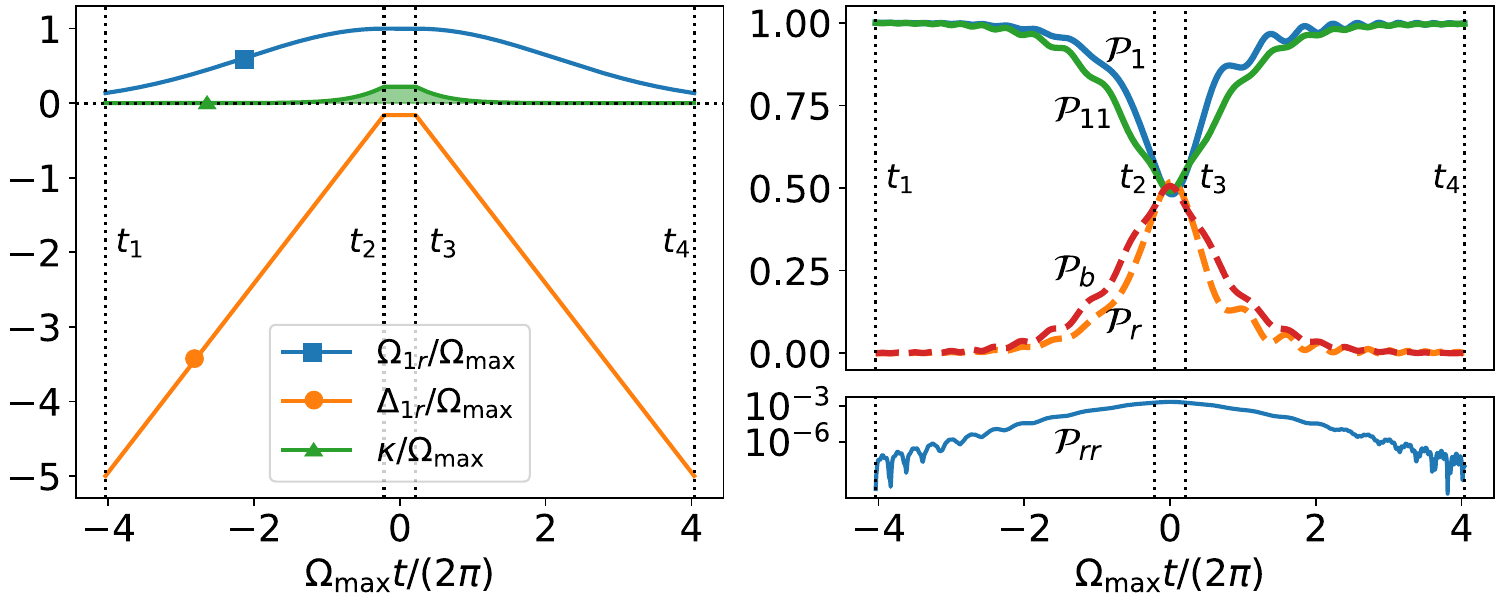}
    \caption{Adiabatic ramp to implement a unitary transformation in \cref{eq:OneAxisTwistRotateUnitary} with $\vartheta_2 = \pi /2$, involving dressing from time $t_1$ to $t_2$, holding the parameters constant for a short interval $t_2$ to $t_3$, and undressing from $t_3$ to $t_4$. Left: Rabi frequency (blue, square); detuning (orange, circle), with a minimum value of $\Delta_{\min}/\Omega_{\max} \approx 0.1 $; and entangling energy (green, triangle) as functions of time during the ramp, with a maximum value of $\kappa_{\max}/\Omega_{\max} \approx 0.25$. Right top: population of $\ket{01}$ and $\ket{10}$ ($\mathcal{P}_1$, blue); population of $\ket{0r}$ and $\ket{r0}$, ($\mathcal{P}_r$ orange, dotted); population of $\ket{11}$ ($\mathcal{P}_{11}$, green); and population of $\ket{b}$ ($\mathcal{P}_b$, red, dotted). Right bottom: population of $\ket{rr}$ ($\mathcal{P}_{rr}$) in a logarithmic scale.
    }
    \label{fig:AdiabaticRamp}
\end{figure}

To implement a two-qubit gate of the form \cref{eq:OneAxisTwistRotateUnitary} we consider an adiabatic ramp in which we sweep both the Rabi frequency $\Omega_{1r}$ and detuning $\Delta_{1r}$ to dress ground states with Rydberg character and then undress them. This implements a rapid adiabatic passage of the logical $|{11}\rangle$ state to a near equal superposition of $|{11}\rangle$ and the entangled bright state $|{b}\rangle$. For a short time we hold the system in this superposition and then perform rapid adiabatic passage back to $|{11}\rangle$. All other logical states involve only single atom dressing or no dressing. As an example, we consider an adiabatic schedule of dressing through a Gaussian ramp of $\Omega_{1r}$ and a linear ramp of the detuning $\Delta_{1r}$ according to
\begin{align}
\qty|\Delta_{1r}(t)|
= \Delta_{\max}
+\frac{ (\Delta_{\max} - \Delta_{\min}) }
   {(t_2 - t_1)}
\qty(t - t_1),
\\ 
\Omega_{1r}(t)
= \Omega_{\min}
+  (\Omega_{\max} - \Omega_{\min})
\exp\qty( - \frac{(t - t_2)^2}{2t_{w}^2}).
\end{align}
After a constant period, we reverse the ramp as shown in \cref{fig:AdiabaticRamp}. The ramp is optimized to achieve a particular value of $\vartheta_2$. 

To implement the MS gate, we consider two adiabatic ramps, each achieving an entangling phase of $\vartheta_2 = \pm \pi/2$, with an echo pulse on the qubit transition, $\exp\left(-\mathrm{i} \pi \hat{S}_x \right)$ between them~\cite{martin2018cphase}. The echo pulse cancels the $\vartheta_1$ accumulated in the two adiabatic ramps, thus implementing a MS gate about the $z$ axis. We convert this to a MS gate about the $y$ axis using $\pi/2$ rotations about the $x$ axis. An advantage of using these adiabatic ramps is they can be designed for any value of $\vartheta_2$, not just integer multiples of $\pi$ as in the pulse sequence $\pi_c - 2\pi_t - \pi_c$ on control $(c)$ and target $(t)$ qubits, proposed in the seminal work of Jaksch \textit{et. al.} ~\cite{jaksch2000fast}. The duration of this ramp, implementing $\vartheta_2 = \pi/2$, shown in  \cref{fig:AdiabaticRamp} is $\approx 8.4 \times 2\pi/\Omega_{\max}$. We calculate the time spent in Rydberg states as the integrated time weighted by the Rydberg population, $t_r = \int \dd t' \mathcal{P}_r(t')$. We find $t_r \approx 0.7 \times 2\pi/\Omega_{\max}$ for initial states $\ket{01}$ and $\ket{10}$ and $t_r \approx 0.9 \times 2\pi/\Omega_{\max}$ for initial state $\ket{11}$. As long as the Rabi period $2\pi/\Omega_{\max}$ is much larger than the radiative lifetime of the Rydberg states, these ramps are rapid and have small loss due to Rydberg state decay. Starting in $|{11}\rangle$ leads to time spent in the doubly excited Rydberg state $|{rr}\rangle$ of $0.0029 \times2\pi/\Omega_{\max}$ when the EDDI is a modest, e.g., $V_{\text{DD}} =10\, \Omega_{\max}$ for $\Omega_{\max}$ of a few  MHz. The population dynamics, during the ramp are shown in \cref{fig:AdiabaticRamp}.

\begin{figure}[t!]
    \centering
    \includegraphics[width=0.48\textwidth]{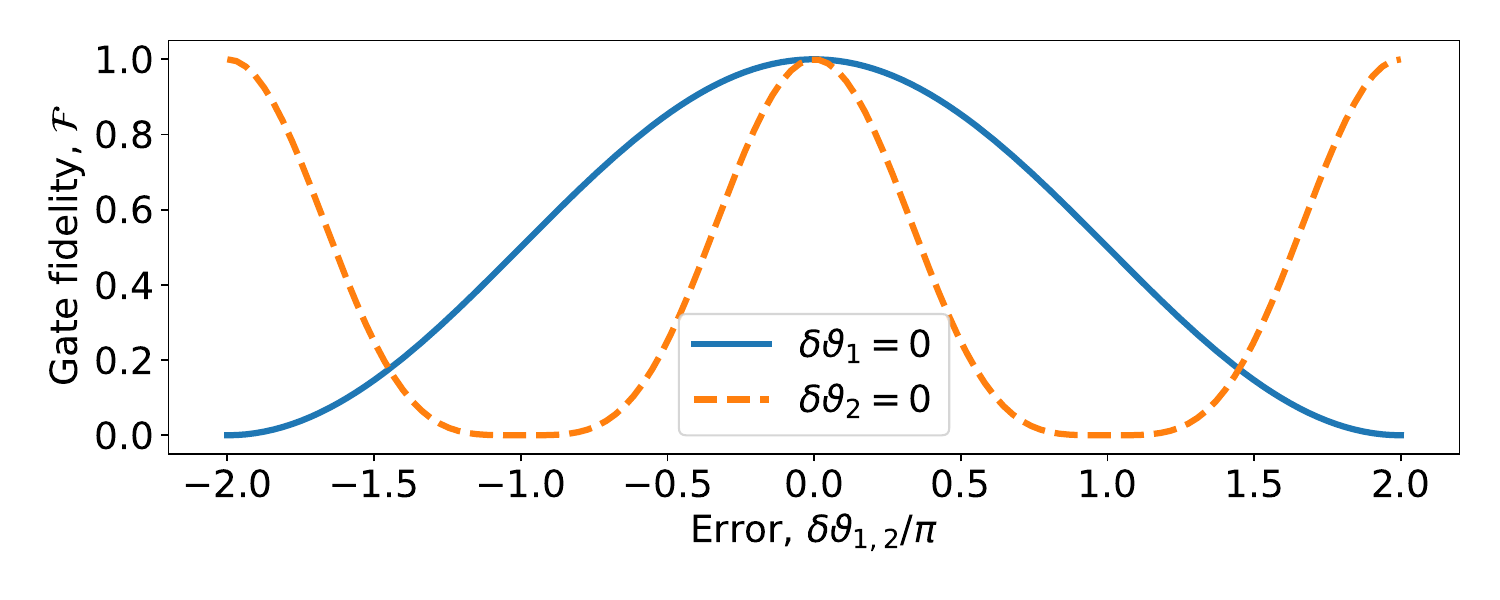}
    \caption{
     Fidelity between a target and implemented unitary transformation, each of the form given in (\cref{eq:OneAxisTwistRotateUnitary}), as a function of the error in the single-qubit rotation angle $\delta \vartheta_1$ (blue, solid line) and as function of error in the two-qubit twist angle $\delta \vartheta_2$ (orange, dotted line). The implementation of the CZ gate is very sensitive to errors caused by inhomogeneities, as it is dominated by $\delta \vartheta_1$, (orange, dotted line).  In contrast, the MS gate is more robust, as it is prone to only errors in $\delta \vartheta_2$ (blue, solid line).
    }
    \label{fig:UnitaryTransformations}
\end{figure}

We assess the performance of the gate by considering the fidelity between the implemented two-qubit gate $\hat{U}$ and the target ideal unitary transformation $\hat{V}$ defined using a normalized Hilbert Schmidt inner product between them $\mathcal{F} =  \qty|\tr\qty(\hat{U} \hat{V}^\dagger)|^2 / 16$, which estimates how well is any input basis is mapped to the corresponding target output basis, by the implemented unitary ~\cite{pedersen2007fidelity}. In particular, we consider errors that can arise from inhomogeneities or coherent errors in the accumulated phases. The fidelity depends on the difference between twist angles $\delta \vartheta_2$ and the difference between rotation angles $\delta \vartheta_1$ of the implemented and target unitary maps according to
\begin{align}
\label{eq:FidelityOneAxistTwistRotate}
    \mathcal{F}
    = \frac{1}{4}
    \qty(1 + \cos^2\qty(\delta \vartheta_1)
    + 2 \cos\qty(\delta \vartheta_1)
    \cos\qty(\frac{\delta \vartheta_2}{2})).
\end{align}
Importantly, the fidelity is much more sensitive to $\delta \vartheta_1$ than it is to $\delta \vartheta_2$. The twist angle $\vartheta_2$ depends solely on the entangling energy $\kappa$. As this is the {\em difference} of two light shifts, it has some common mode cancellation of errors in the light shifts, while $\vartheta_1$ has a contribution from independent single-atom light shifts with no such cancellation. This effect is seen in \cref{fig:UnitaryTransformations} which shows the fidelity plotted as a function of $\delta \vartheta_1$ when $\delta \vartheta_2 = 0$, that is, $\mathcal{F} = \tfrac{1}{2} \qty(1 + \cos\qty(\delta \vartheta_2 / 2))$ and as a function of $\delta \vartheta_2$ when $\delta \vartheta_1 = 0$, that is, $\mathcal{F} = \tfrac{1}{4} \qty(1 + \cos\qty(\delta \vartheta_1))^2$. Note, the CZ gate studied in~\cite{keating2015robust} required knowledge of $E_{\text{LS}}^{(1)}$ to remove the single-atom contribution to the phase, and errors will contribute substantially to infidelity through $\delta \vartheta_1$. In contrast, the MS gate is substantially less sensitive to such errors, as  $\delta \vartheta_1$ can be made zero by using a spin echo~\cite{martin2018cphase}.


Let us consider the error channels and the intrinsic robustness of using adiabatic Rydberg dressing to implement the MS gate. Deleterious effects include thermal Doppler shifts and atomic motion in a spatially inhomogeneous exciting laser, imperfect blockade, and  finite radiative lifetime of the Rydberg state. To see how this effect arises, let us revisit the dressed states, including the quantized motion. For generality we include quantized atomic momenta $p_\alpha$ and $p_\beta$ of the two atoms in their Rydberg dressing interaction in addition to the electronic ground state and the bright and dark states. The bare states are the ground state $\ket{G} = \ket{1, p_\alpha; 1, p_\beta}$, bright state $\ket{B} = \tfrac{1}{\sqrt{2}}
\left( \ket{r, p_\alpha + k_{1r}; 1, p_\beta} + \ket{1, p_\alpha; r, p_\beta + k_{1r}}\right)
$ and dark state $\ket{D}= \tfrac{1}{\sqrt{2}}
\left(\ket{r, p_\alpha + k_{1r}; 1, p_\beta} - \ket{1, p_\alpha; r, p_\beta + k_{1r}}\right)$. The two-atom Rydberg Hamiltonian now generalizes to ~\cite{keating2015robust}
\begin{widetext}

\begin{multline}
    \hat{H}_{\text{2atom}}(p_\alpha,p_\beta)=
    -\left(\Delta_{1r}-\frac{k_{1r} P_{\mathrm{CM}}}{M}\right)
    \left(|{B}\rangle\langle{B}| + |{D}\rangle\langle{D}|\right)
    + \left(V_{\text{DD}} -2\left(\Delta_{1r}-\frac{k_{1r} P_{\mathrm{CM}}}{M}\right)\right) 
    |{rr}\rangle\langle{rr}|
    \\
    + \frac{k_{1r} p_{\mathrm{rel}}}{m}
    \left(|{B}\rangle\langle{D}| +
    \text{h.c.} \right) 
    + \frac{1}{2} \left(\frac{\Omega_\alpha + \Omega_\beta}{\sqrt{2}}\right)
    \left(|{B}\rangle\langle{G}| + 
    |{rr}\rangle\langle{B}| +
    \text{h.c.} \right) 
    + \frac{1}{2} \left(\frac{\Omega_\alpha - \Omega_\beta}{\sqrt{2}}\right)
    \left(|{D}\rangle\langle{G}| +
    |{rr}\rangle\langle{D}| +
    \text{h.c.} \right)
\end{multline}

\end{widetext}
where $\Omega_\alpha = \Omega_{1r}(z_\alpha)$ and $\Omega_\beta = \Omega_{1r}(z_\beta)$ are the Rabi frequencies at the positions of atoms $\alpha$ and $\beta$; $P_{\mathrm{CM}} = p_\alpha+p_\beta$ and $p_{\mathrm{rel}} = (p_\alpha-p_\beta)/2$ are the center-of-mass and relative momenta of the atoms ~\cite{keating2015robust}.

\begin{figure*}
    \centering
     \includegraphics[width=0.96\textwidth]
    {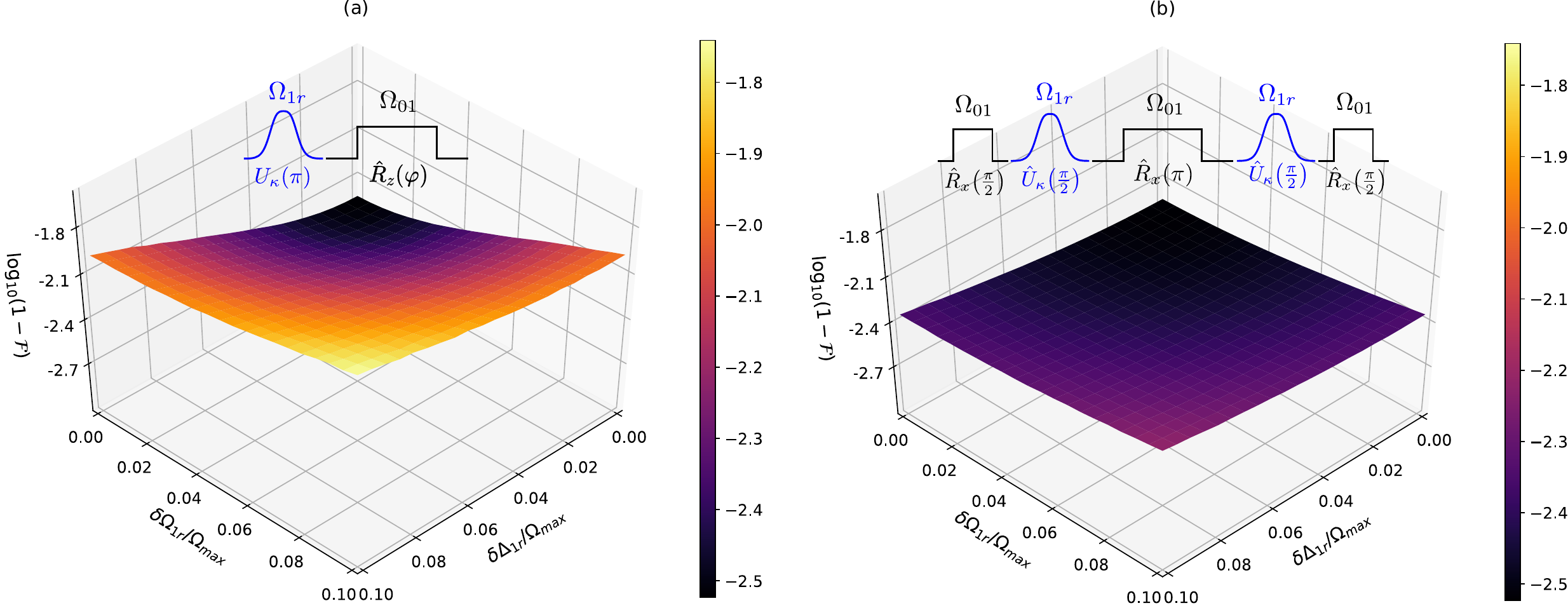}
     \caption{(a) Top: Implementing the CZ gate as proposed in ~\cite{keating2015robust} using an adiabatic ramp, followed by removal of phases accumulated due to one atom light shifts using a single qubit rotation $\hat{R}_z(\varphi)$, where $\varphi = \int \dd t' E_{\text{LS}}^{(1)}(t')$.
    Bottom: Simulated infidelities of implementing the CZ gate with different levels of inhomogeneities in $\Delta_{1r}$ and $\Omega_{1r}$.
    (b) Top: Implementing the MS gate as done in ~\cite{martin2018cphase} using two adiabatic ramps, with an spin echo in between.
    Bottom: Simulated infidelities of implementing the MS gate with different levels of inhomogeneities in $\Delta_{1r}$ and $\Omega_{1r}$.
    }
    \label{fig:Protocols}
\end{figure*}

The standard protocol of Jaksch {\em et al.}~\cite{jaksch2000fast} involves a pulse sequence on the control($c$) and target($t$) qubits, $\pi_c - 2\pi_t - \pi_c$, ideally yielding a CZ gate. In the presence of thermal atomic velocity $v_c$ for the control atom, the transformation on the logical states is $\ket{00}\rightarrow \ket{00}$, $\ket{01}\rightarrow -\ket{01}$, $\ket{10}\rightarrow -\mathrm{e}^{-\mathrm{i} k_{1r} v_c\delta t}\ket{10}$, $\ket{11}\rightarrow -\mathrm{e}^{-\mathrm{i} k_{1r} v_c \delta t}\ket{11}$. Relative to the ideal CZ gate, there are additional phases due to the random Doppler shift acquired when the control atom stays in the Rydberg state for a time $\delta t$. For a thermal distribution of momenta, the random distribution of phases cannot be compensated, which causes gate errors ~\cite{wilk2010entanglement, keating2015robust, levine2018high, levine2019parallel, graham2019rydberg}. In contrast to the direct excitation to Rydberg states, for adiabatic Rydberg dressing, there are no random phases imparted to the qubits. Instead, the center-of-mass motion leads to a detuning error ~\cite{keating2015robust}. The relative motion leads to coupling between bright and dark states. However, while using an adiabatic ramp, this is suppressed due to the energy gap between the light-shifted bright state and unshifted dark state. The residual off-resonance $\ket*{\tilde{B}} \leftrightarrow \ket{D}$ coupling leads to a small second order perturbative shift on the dressed ground state ~\cite{keating2015robust}. Moreover, a nonuniform intensity in which atoms see different Rabi frequencies can introduce a coupling between the ground $\ket{11}$ and the dark state $\ket{D}$, which gives a small perturbative shift on the dressed ground state.

Finally there is the effect of imperfect blockade. Whereas in the standard pulsed protocol this can be a major source of error, gates based on adiabatic dressing are more resilient to this effect. As long as the evolution is adiabatic, the dressed ground states will contain a small admixture of doubly-excited Rydberg states in the superposition as shown in \cref{fig:AdiabaticRamp}. This will affect the value of $\kappa$, but this can be measured, and the adiabatic ramp can be adjusted accordingly. If we are close to the blockade radius, the gradient of dressed ground state energy as a function of separation between the atoms will be small, and there will be negligible force on the atoms due to the EDDI. Of course non-adiabatic effects such as resonant excitation to other doubly-excited Rydberg states can add additional errors, but these effects are not studied here.

As an example we consider ${}^{133}\mathrm{Cs}$ as used in our experiment, with qubit states  $\ket{1} = |{6 \mathrm{S}_{1/2}, F=4, m_F=0}\rangle$ and $\ket{0} = \ket{6 \mathrm{S}_{1/2}, F=3, m_F=0}$ \cite{jau2016entangling, lee2017demonstration}, with $\ket{1}$ coupled through a one-photon transition at $319\,\mathrm{nm}$ (the ``Rydberg laser") to state $\ket{r} = \ket{64 \mathrm{P}_{3/2},  m_J=3/2}$. The qubit states can be connected with a microwave field or Raman laser fields. A maximum Rydberg Rabi frequency of $\Omega_{\mathrm{max}} / 2\pi = 4$  MHz, gives an entangling strength at our final detuning near resonance ($\Delta_{\min} \approx 0.1 \Omega_{\max}$) of $\kappa / 2\pi \approx 1$, MHz under the perfect blockade approximation. A contribution to detuning inhomogeneities arises from the width of atomic velocity distribution, $\propto \sqrt{k_{\mathrm{B}} T / m}$ ~\cite{wilk2010entanglement, zhang2012fidelity, walker2012entanglement, de2018analysis, graham2019rydberg}. The width of the distribution of $\kappa$ due to a thermal Doppler width of the atomic momenta distribution can be estimated as $    \delta \kappa_{\text{th}} = k_{1r} \sqrt{\frac{k_{\mathrm{B}} T}{m}}  \sqrt{\qty(\pdv{\kappa}{\Delta_\alpha})^2 + \qty(\pdv{\kappa}{\Delta_\beta})^2}$, where $\Delta_\alpha = \Delta_{1r} - k_{1r} p_{\alpha} / m$ and similarly for $\Delta_\beta$. For our experiment, an atomic temperature of $10$ $\mathrm{\mu K}$ \cite{lee2017demonstration} corresponds to $k_{1r} \sqrt{k_{\mathrm{B}} T /m} \approx  0.02 \Omega_{\max}$, therefore $\delta \kappa \sim 10^{-2} \Omega_{\max}$, leading to $\delta \vartheta_2/ \vartheta_2 \approx 4\times 10^{-2}$ for the example parameters used here. The Rydberg laser uses a Gaussian beam with a waist of about $ 15\,\mathrm{\mu m}$, the transverse atomic position spread is about $1.45\,\mathrm{\mu m}$ and the atomic separation is about $2\,\mathrm{\mu m}$. This gives us a Rabi frequency inhomogeneity of $\delta \Omega_{1r}  \approx 0.01 \Omega_{1r}$.

We model the experimental scenario by considering the detuning $\Delta_{1r}$ and Rabi frequency $\Omega_{1r}$ for each atom to be sampled from a normal distribution with mean equal to the fiducial value and standard deviation determined by the level of imperfections in the experiment. We simulate the implementation of the CZ gate using the protocol proposed earlier ~\cite{keating2015robust} and the implementation of the MS gate using two adiabatic ramps and a spin echo, over a range of inhomogeneities $\delta \Delta_{1r}$ and $\delta \Omega_{1r}$. The gate fidelity including inhomogeneities, imperfect blockade, and Rydberg state decay (Rydberg lifetime $\tau_r \sim 10^3 (2\pi/\Omega_{\max})$) for an EDDI strength of $10 \, \Omega_{\max}$ and the target gate is shown in \cref{fig:Protocols} (a) for the CZ gate \cref{fig:Protocols} (b) for the MS gate. As expected from \cref{fig:UnitaryTransformations}, we see that implementing the MS gate using two adiabatic ramps and a spin echo is much more robust to inhomogeneities in $\Omega_{1r}$ and $\Delta_{1r}$ than implementing the CZ gate using an adiabatic ramp (\cref{fig:Protocols}). For example, when we increase the level of imperfections from $0$ to about $10\%$ of the maximum Rabi frequency $\Omega_{\max}$ in the Rabi frequency and detuning, the MS gate fidelity falls from about $0.997$ to about $0.995$, while the CZ gate fidelity falls from about $0.997$ to about $0.986$. Therefore, high-fidelity gates are possible at room temperature with modest radiative lifetimes, consistent with the demands of adiabatic evolution.

Ultimately, the best achievable adiabaticity is limited by the finite radiative lifetime of the Rydberg states, which is the fundamental source of error \cite{saffman2010quantum, saffman2016quantum} and limits the adiabaticity of the adiabatic ramp. The effective lifetime $\tau_r$ is due to contributions of different decay channels including ionization, spontaneous emission, and stimulated emission via coupling to blackbody radiation. The effect of finite radiative lifetime of the Rydberg state can be estimated by the quantity $\pi / (\kappa \tau_r) $ which compares how quickly atoms decay versus how quickly they accumulate entangling phase. The entangling phase is accumulated faster with a larger Rabi frequency, $\Omega_{1r}$. More precisely, the time spent by atoms in the Rydberg state is quantified by the integrated Rydberg population, described above. Optimizing the parameters of the adiabatic ramp, we can satisfy $t_r / \tau_r \ll 1$ as long as $\Omega_{\max} \tau_r \ll 1$. For example, for a lifetime of $\tau_r = 140 \mathrm{\mu s}$ \cite{beterov2009quasiclassical, levine2018high} and given $\kappa / 2\pi \approx 1\, \mathrm{MHz}$,  $\pi / (\kappa \tau_r) \sim 10^{-3}$. Lifetimes of a few milliseconds, achievable by choosing higher lying Rydberg states and for cryogenic environments of a few Kelvin~\cite{saffman2016quantum} would give $\pi / (\kappa \tau_r) \sim 10^{-4}$.


In summary, adiabatic Rydberg dressing provides a robust method for harnessing the EDDI between Rydberg excited atoms to generate entanglement between qubits encoded in atomic clock states. We have shown that with current experimental capabilities, a two qubit MS entangling gate with a fidelity of $\sim 0.995$ is within reach by interleaving of adiabatic Rydberg dressing and undressing with a spin echo on the qubit transition. Even higher fidelity gates are possible at cryogenic temperatures which substantially increases the Rydberg state lifetime. Such longer coherence times allow for improved adiabatic ramps and the potential use of more sophisticated robust control~\cite{goerz2014robustness} to correct residual inhomogeneities not canceled in simple spin echo.

\begin{acknowledgments}
We thank Christiane Koch and David Weiss for helpful discussions. This work was supported by NSF grants PHY-1606989, PHY-1630114, the Laboratory Directed Research and Development program at Sandia National Laboratories, and by the Laboratory Directed Research and Development program of Los Alamos National Laboratory under project numbers 20190494ER and 20200015ER. Sandia National Laboratories is a multimission laboratory managed and operated by National Technology \& Engineering Solutions of Sandia, LLC, a wholly owned subsidiary of Honeywell International Inc., for the U.S. Department of Energy National Nuclear Security Administration under contract DE-NA0003525. This paper describes objective technical results and analysis. Any subjective views or opinions that might be expressed in the paper do not necessarily represent the views of the U.S. Department of Energy or the United States Government.
\end{acknowledgments}

\bibliography{references}

\begin{thebibliography}{58}%
\makeatletter
\providecommand \@ifxundefined [1]{%
 \@ifx{#1\undefined}
}%
\providecommand \@ifnum [1]{%
 \ifnum #1\expandafter \@firstoftwo
 \else \expandafter \@secondoftwo
 \fi
}%
\providecommand \@ifx [1]{%
 \ifx #1\expandafter \@firstoftwo
 \else \expandafter \@secondoftwo
 \fi
}%
\providecommand \natexlab [1]{#1}%
\providecommand \enquote  [1]{``#1''}%
\providecommand \bibnamefont  [1]{#1}%
\providecommand \bibfnamefont [1]{#1}%
\providecommand \citenamefont [1]{#1}%
\providecommand \href@noop [0]{\@secondoftwo}%
\providecommand \href [0]{\begingroup \@sanitize@url \@href}%
\providecommand \@href[1]{\@@startlink{#1}\@@href}%
\providecommand \@@href[1]{\endgroup#1\@@endlink}%
\providecommand \@sanitize@url [0]{\catcode `\\12\catcode `\$12\catcode
  `\&12\catcode `\#12\catcode `\^12\catcode `\_12\catcode `\%12\relax}%
\providecommand \@@startlink[1]{}%
\providecommand \@@endlink[0]{}%
\providecommand \url  [0]{\begingroup\@sanitize@url \@url }%
\providecommand \@url [1]{\endgroup\@href {#1}{\urlprefix }}%
\providecommand \urlprefix  [0]{URL }%
\providecommand \Eprint [0]{\href }%
\providecommand \doibase [0]{https://doi.org/}%
\providecommand \selectlanguage [0]{\@gobble}%
\providecommand \bibinfo  [0]{\@secondoftwo}%
\providecommand \bibfield  [0]{\@secondoftwo}%
\providecommand \translation [1]{[#1]}%
\providecommand \BibitemOpen [0]{}%
\providecommand \bibitemStop [0]{}%
\providecommand \bibitemNoStop [0]{.\EOS\space}%
\providecommand \EOS [0]{\spacefactor3000\relax}%
\providecommand \BibitemShut  [1]{\csname bibitem#1\endcsname}%
\let\auto@bib@innerbib\@empty
\bibitem [{\citenamefont {Brennen}\ \emph {et~al.}(1999)\citenamefont
  {Brennen}, \citenamefont {Caves}, \citenamefont {Jessen},\ and\ \citenamefont
  {Deutsch}}]{brennen1999quantum}%
  \BibitemOpen
  \bibfield  {author} {\bibinfo {author} {\bibfnamefont {G.~K.}\ \bibnamefont
  {Brennen}}, \bibinfo {author} {\bibfnamefont {C.~M.}\ \bibnamefont {Caves}},
  \bibinfo {author} {\bibfnamefont {P.~S.}\ \bibnamefont {Jessen}},\ and\
  \bibinfo {author} {\bibfnamefont {I.~H.}\ \bibnamefont {Deutsch}},\
  }\href@noop {} {\bibfield  {journal} {\bibinfo  {journal} {Phys. Rev. Lett.}\
  }\textbf {\bibinfo {volume} {82}},\ \bibinfo {pages} {1060} (\bibinfo {year}
  {1999})}\BibitemShut {NoStop}%
\bibitem [{\citenamefont {Brennen}\ \emph {et~al.}(2000)\citenamefont
  {Brennen}, \citenamefont {Deutsch},\ and\ \citenamefont
  {Jessen}}]{brennen2000entangling}%
  \BibitemOpen
  \bibfield  {author} {\bibinfo {author} {\bibfnamefont {G.~K.}\ \bibnamefont
  {Brennen}}, \bibinfo {author} {\bibfnamefont {I.~H.}\ \bibnamefont
  {Deutsch}},\ and\ \bibinfo {author} {\bibfnamefont {P.~S.}\ \bibnamefont
  {Jessen}},\ }\href@noop {} {\bibfield  {journal} {\bibinfo  {journal} {Phys.
  Rev. A}\ }\textbf {\bibinfo {volume} {61}},\ \bibinfo {pages} {062309}
  (\bibinfo {year} {2000})}\BibitemShut {NoStop}%
\bibitem [{\citenamefont {Weiss}\ and\ \citenamefont
  {Saffman}(2017)}]{weiss2017quantum}%
  \BibitemOpen
  \bibfield  {author} {\bibinfo {author} {\bibfnamefont {D.~S.}\ \bibnamefont
  {Weiss}}\ and\ \bibinfo {author} {\bibfnamefont {M.}~\bibnamefont
  {Saffman}},\ }\href@noop {} {\bibfield  {journal} {\bibinfo  {journal}
  {Physics Today}\ }\textbf {\bibinfo {volume} {70}} (\bibinfo {year}
  {2017})}\BibitemShut {NoStop}%
\bibitem [{\citenamefont {Labuhn}\ \emph {et~al.}(2016)\citenamefont {Labuhn},
  \citenamefont {Barredo}, \citenamefont {Ravets}, \citenamefont
  {De~L{\'e}s{\'e}leuc}, \citenamefont {Macr{\`\i}}, \citenamefont {Lahaye},\
  and\ \citenamefont {Browaeys}}]{labuhn2016tunable}%
  \BibitemOpen
  \bibfield  {author} {\bibinfo {author} {\bibfnamefont {H.}~\bibnamefont
  {Labuhn}}, \bibinfo {author} {\bibfnamefont {D.}~\bibnamefont {Barredo}},
  \bibinfo {author} {\bibfnamefont {S.}~\bibnamefont {Ravets}}, \bibinfo
  {author} {\bibfnamefont {S.}~\bibnamefont {De~L{\'e}s{\'e}leuc}}, \bibinfo
  {author} {\bibfnamefont {T.}~\bibnamefont {Macr{\`\i}}}, \bibinfo {author}
  {\bibfnamefont {T.}~\bibnamefont {Lahaye}},\ and\ \bibinfo {author}
  {\bibfnamefont {A.}~\bibnamefont {Browaeys}},\ }\href@noop {} {\bibfield
  {journal} {\bibinfo  {journal} {Nature}\ }\textbf {\bibinfo {volume} {534}},\
  \bibinfo {pages} {667} (\bibinfo {year} {2016})}\BibitemShut {NoStop}%
\bibitem [{\citenamefont {de~L{\'e}s{\'e}leuc}\ \emph
  {et~al.}(2018{\natexlab{a}})\citenamefont {de~L{\'e}s{\'e}leuc},
  \citenamefont {Weber}, \citenamefont {Lienhard}, \citenamefont {Barredo},
  \citenamefont {B{\"u}chler}, \citenamefont {Lahaye},\ and\ \citenamefont
  {Browaeys}}]{de2018accurate}%
  \BibitemOpen
  \bibfield  {author} {\bibinfo {author} {\bibfnamefont {S.}~\bibnamefont
  {de~L{\'e}s{\'e}leuc}}, \bibinfo {author} {\bibfnamefont {S.}~\bibnamefont
  {Weber}}, \bibinfo {author} {\bibfnamefont {V.}~\bibnamefont {Lienhard}},
  \bibinfo {author} {\bibfnamefont {D.}~\bibnamefont {Barredo}}, \bibinfo
  {author} {\bibfnamefont {H.~P.}\ \bibnamefont {B{\"u}chler}}, \bibinfo
  {author} {\bibfnamefont {T.}~\bibnamefont {Lahaye}},\ and\ \bibinfo {author}
  {\bibfnamefont {A.}~\bibnamefont {Browaeys}},\ }\href@noop {} {\bibfield
  {journal} {\bibinfo  {journal} {Phys. Rev. Lett.}\ }\textbf {\bibinfo
  {volume} {120}},\ \bibinfo {pages} {113602} (\bibinfo {year}
  {2018}{\natexlab{a}})}\BibitemShut {NoStop}%
\bibitem [{\citenamefont {Keating}\ \emph {et~al.}(2013)\citenamefont
  {Keating}, \citenamefont {Goyal}, \citenamefont {Jau}, \citenamefont
  {Biedermann}, \citenamefont {Landahl},\ and\ \citenamefont
  {Deutsch}}]{keating2013adiabatic}%
  \BibitemOpen
  \bibfield  {author} {\bibinfo {author} {\bibfnamefont {T.}~\bibnamefont
  {Keating}}, \bibinfo {author} {\bibfnamefont {K.}~\bibnamefont {Goyal}},
  \bibinfo {author} {\bibfnamefont {Y.-Y.}\ \bibnamefont {Jau}}, \bibinfo
  {author} {\bibfnamefont {G.~W.}\ \bibnamefont {Biedermann}}, \bibinfo
  {author} {\bibfnamefont {A.~J.}\ \bibnamefont {Landahl}},\ and\ \bibinfo
  {author} {\bibfnamefont {I.~H.}\ \bibnamefont {Deutsch}},\ }\href@noop {}
  {\bibfield  {journal} {\bibinfo  {journal} {Phys. Rev. A}\ }\textbf {\bibinfo
  {volume} {87}},\ \bibinfo {pages} {052314} (\bibinfo {year}
  {2013})}\BibitemShut {NoStop}%
\bibitem [{\citenamefont {Zhou}\ \emph {et~al.}(2018)\citenamefont {Zhou},
  \citenamefont {Wang}, \citenamefont {Choi}, \citenamefont {Pichler},\ and\
  \citenamefont {Lukin}}]{zhou2018quantum}%
  \BibitemOpen
  \bibfield  {author} {\bibinfo {author} {\bibfnamefont {L.}~\bibnamefont
  {Zhou}}, \bibinfo {author} {\bibfnamefont {S.-T.}\ \bibnamefont {Wang}},
  \bibinfo {author} {\bibfnamefont {S.}~\bibnamefont {Choi}}, \bibinfo {author}
  {\bibfnamefont {H.}~\bibnamefont {Pichler}},\ and\ \bibinfo {author}
  {\bibfnamefont {M.~D.}\ \bibnamefont {Lukin}},\ }\href@noop {} {\bibfield
  {journal} {\bibinfo  {journal} {arXiv preprint arXiv:1812.01041}\ } (\bibinfo
  {year} {2018})}\BibitemShut {NoStop}%
\bibitem [{\citenamefont {Pichler}\ \emph {et~al.}(2018)\citenamefont
  {Pichler}, \citenamefont {Wang}, \citenamefont {Zhou}, \citenamefont {Choi},\
  and\ \citenamefont {Lukin}}]{pichler2018computational}%
  \BibitemOpen
  \bibfield  {author} {\bibinfo {author} {\bibfnamefont {H.}~\bibnamefont
  {Pichler}}, \bibinfo {author} {\bibfnamefont {S.-T.}\ \bibnamefont {Wang}},
  \bibinfo {author} {\bibfnamefont {L.}~\bibnamefont {Zhou}}, \bibinfo {author}
  {\bibfnamefont {S.}~\bibnamefont {Choi}},\ and\ \bibinfo {author}
  {\bibfnamefont {M.~D.}\ \bibnamefont {Lukin}},\ }\href@noop {} {\bibfield
  {journal} {\bibinfo  {journal} {arXiv preprint arXiv:1809.04954}\ } (\bibinfo
  {year} {2018})}\BibitemShut {NoStop}%
\bibitem [{\citenamefont {Ludlow}\ \emph {et~al.}(2015)\citenamefont {Ludlow},
  \citenamefont {Boyd}, \citenamefont {Ye}, \citenamefont {Peik},\ and\
  \citenamefont {Schmidt}}]{ludlow2015optical}%
  \BibitemOpen
  \bibfield  {author} {\bibinfo {author} {\bibfnamefont {A.~D.}\ \bibnamefont
  {Ludlow}}, \bibinfo {author} {\bibfnamefont {M.~M.}\ \bibnamefont {Boyd}},
  \bibinfo {author} {\bibfnamefont {J.}~\bibnamefont {Ye}}, \bibinfo {author}
  {\bibfnamefont {E.}~\bibnamefont {Peik}},\ and\ \bibinfo {author}
  {\bibfnamefont {P.~O.}\ \bibnamefont {Schmidt}},\ }\href@noop {} {\bibfield
  {journal} {\bibinfo  {journal} {Rev. Mod. Phys.}\ }\textbf {\bibinfo {volume}
  {87}},\ \bibinfo {pages} {637} (\bibinfo {year} {2015})}\BibitemShut
  {NoStop}%
\bibitem [{\citenamefont {Endres}\ \emph {et~al.}(2016)\citenamefont {Endres},
  \citenamefont {Bernien}, \citenamefont {Keesling}, \citenamefont {Levine},
  \citenamefont {Anschuetz}, \citenamefont {Krajenbrink}, \citenamefont
  {Senko}, \citenamefont {Vuletic}, \citenamefont {Greiner},\ and\
  \citenamefont {Lukin}}]{endres2016atom}%
  \BibitemOpen
  \bibfield  {author} {\bibinfo {author} {\bibfnamefont {M.}~\bibnamefont
  {Endres}}, \bibinfo {author} {\bibfnamefont {H.}~\bibnamefont {Bernien}},
  \bibinfo {author} {\bibfnamefont {A.}~\bibnamefont {Keesling}}, \bibinfo
  {author} {\bibfnamefont {H.}~\bibnamefont {Levine}}, \bibinfo {author}
  {\bibfnamefont {E.~R.}\ \bibnamefont {Anschuetz}}, \bibinfo {author}
  {\bibfnamefont {A.}~\bibnamefont {Krajenbrink}}, \bibinfo {author}
  {\bibfnamefont {C.}~\bibnamefont {Senko}}, \bibinfo {author} {\bibfnamefont
  {V.}~\bibnamefont {Vuletic}}, \bibinfo {author} {\bibfnamefont
  {M.}~\bibnamefont {Greiner}},\ and\ \bibinfo {author} {\bibfnamefont {M.~D.}\
  \bibnamefont {Lukin}},\ }\href@noop {} {\bibfield  {journal} {\bibinfo
  {journal} {Science}\ }\textbf {\bibinfo {volume} {354}},\ \bibinfo {pages}
  {1024} (\bibinfo {year} {2016})}\BibitemShut {NoStop}%
\bibitem [{\citenamefont {Barredo}\ \emph {et~al.}(2016)\citenamefont
  {Barredo}, \citenamefont {De~L{\'e}s{\'e}leuc}, \citenamefont {Lienhard},
  \citenamefont {Lahaye},\ and\ \citenamefont {Browaeys}}]{barredo2016atom}%
  \BibitemOpen
  \bibfield  {author} {\bibinfo {author} {\bibfnamefont {D.}~\bibnamefont
  {Barredo}}, \bibinfo {author} {\bibfnamefont {S.}~\bibnamefont
  {De~L{\'e}s{\'e}leuc}}, \bibinfo {author} {\bibfnamefont {V.}~\bibnamefont
  {Lienhard}}, \bibinfo {author} {\bibfnamefont {T.}~\bibnamefont {Lahaye}},\
  and\ \bibinfo {author} {\bibfnamefont {A.}~\bibnamefont {Browaeys}},\
  }\href@noop {} {\bibfield  {journal} {\bibinfo  {journal} {Science}\ }\textbf
  {\bibinfo {volume} {354}},\ \bibinfo {pages} {1021} (\bibinfo {year}
  {2016})}\BibitemShut {NoStop}%
\bibitem [{\citenamefont {Barredo}\ \emph {et~al.}(2018)\citenamefont
  {Barredo}, \citenamefont {Lienhard}, \citenamefont {De~Leseleuc},
  \citenamefont {Lahaye},\ and\ \citenamefont
  {Browaeys}}]{barredo2018synthetic}%
  \BibitemOpen
  \bibfield  {author} {\bibinfo {author} {\bibfnamefont {D.}~\bibnamefont
  {Barredo}}, \bibinfo {author} {\bibfnamefont {V.}~\bibnamefont {Lienhard}},
  \bibinfo {author} {\bibfnamefont {S.}~\bibnamefont {De~Leseleuc}}, \bibinfo
  {author} {\bibfnamefont {T.}~\bibnamefont {Lahaye}},\ and\ \bibinfo {author}
  {\bibfnamefont {A.}~\bibnamefont {Browaeys}},\ }\href@noop {} {\bibfield
  {journal} {\bibinfo  {journal} {Nature}\ }\textbf {\bibinfo {volume} {561}},\
  \bibinfo {pages} {79} (\bibinfo {year} {2018})}\BibitemShut {NoStop}%
\bibitem [{\citenamefont {Kumar}\ \emph {et~al.}(2018)\citenamefont {Kumar},
  \citenamefont {Wu}, \citenamefont {Giraldo},\ and\ \citenamefont
  {Weiss}}]{kumar2018sorting}%
  \BibitemOpen
  \bibfield  {author} {\bibinfo {author} {\bibfnamefont {A.}~\bibnamefont
  {Kumar}}, \bibinfo {author} {\bibfnamefont {T.-Y.}\ \bibnamefont {Wu}},
  \bibinfo {author} {\bibfnamefont {F.}~\bibnamefont {Giraldo}},\ and\ \bibinfo
  {author} {\bibfnamefont {D.~S.}\ \bibnamefont {Weiss}},\ }\href@noop {}
  {\bibfield  {journal} {\bibinfo  {journal} {Nature}\ }\textbf {\bibinfo
  {volume} {561}},\ \bibinfo {pages} {83} (\bibinfo {year} {2018})}\BibitemShut
  {NoStop}%
\bibitem [{\citenamefont {Wang}\ \emph {et~al.}(2015)\citenamefont {Wang},
  \citenamefont {Zhang}, \citenamefont {Corcovilos}, \citenamefont {Kumar},\
  and\ \citenamefont {Weiss}}]{wang2015coherent}%
  \BibitemOpen
  \bibfield  {author} {\bibinfo {author} {\bibfnamefont {Y.}~\bibnamefont
  {Wang}}, \bibinfo {author} {\bibfnamefont {X.}~\bibnamefont {Zhang}},
  \bibinfo {author} {\bibfnamefont {T.~A.}\ \bibnamefont {Corcovilos}},
  \bibinfo {author} {\bibfnamefont {A.}~\bibnamefont {Kumar}},\ and\ \bibinfo
  {author} {\bibfnamefont {D.~S.}\ \bibnamefont {Weiss}},\ }\href@noop {}
  {\bibfield  {journal} {\bibinfo  {journal} {Phys. Rev. Lett.}\ }\textbf
  {\bibinfo {volume} {115}},\ \bibinfo {pages} {043003} (\bibinfo {year}
  {2015})}\BibitemShut {NoStop}%
\bibitem [{\citenamefont {Wang}\ \emph {et~al.}(2016)\citenamefont {Wang},
  \citenamefont {Kumar}, \citenamefont {Wu},\ and\ \citenamefont
  {Weiss}}]{wang2016single}%
  \BibitemOpen
  \bibfield  {author} {\bibinfo {author} {\bibfnamefont {Y.}~\bibnamefont
  {Wang}}, \bibinfo {author} {\bibfnamefont {A.}~\bibnamefont {Kumar}},
  \bibinfo {author} {\bibfnamefont {T.-Y.}\ \bibnamefont {Wu}},\ and\ \bibinfo
  {author} {\bibfnamefont {D.~S.}\ \bibnamefont {Weiss}},\ }\href@noop {}
  {\bibfield  {journal} {\bibinfo  {journal} {Science}\ }\textbf {\bibinfo
  {volume} {352}},\ \bibinfo {pages} {1562} (\bibinfo {year}
  {2016})}\BibitemShut {NoStop}%
\bibitem [{\citenamefont {Jaksch}\ \emph {et~al.}(1999)\citenamefont {Jaksch},
  \citenamefont {Briegel}, \citenamefont {Cirac}, \citenamefont {Gardiner},\
  and\ \citenamefont {Zoller}}]{jaksch1999entanglement}%
  \BibitemOpen
  \bibfield  {author} {\bibinfo {author} {\bibfnamefont {D.}~\bibnamefont
  {Jaksch}}, \bibinfo {author} {\bibfnamefont {H.-J.}\ \bibnamefont {Briegel}},
  \bibinfo {author} {\bibfnamefont {J.-I.}\ \bibnamefont {Cirac}}, \bibinfo
  {author} {\bibfnamefont {C.-W.}\ \bibnamefont {Gardiner}},\ and\ \bibinfo
  {author} {\bibfnamefont {P.}~\bibnamefont {Zoller}},\ }\href@noop {}
  {\bibfield  {journal} {\bibinfo  {journal} {Phys. Rev. Lett.}\ }\textbf
  {\bibinfo {volume} {82}},\ \bibinfo {pages} {1975} (\bibinfo {year}
  {1999})}\BibitemShut {NoStop}%
\bibitem [{\citenamefont {Lukin}\ \emph {et~al.}(2001)\citenamefont {Lukin},
  \citenamefont {Fleischhauer}, \citenamefont {Cote}, \citenamefont {Duan},
  \citenamefont {Jaksch}, \citenamefont {Cirac},\ and\ \citenamefont
  {Zoller}}]{lukin2001dipole}%
  \BibitemOpen
  \bibfield  {author} {\bibinfo {author} {\bibfnamefont {M.-D.}\ \bibnamefont
  {Lukin}}, \bibinfo {author} {\bibfnamefont {M.}~\bibnamefont {Fleischhauer}},
  \bibinfo {author} {\bibfnamefont {R.}~\bibnamefont {Cote}}, \bibinfo {author}
  {\bibfnamefont {L.-M.}\ \bibnamefont {Duan}}, \bibinfo {author}
  {\bibfnamefont {D.}~\bibnamefont {Jaksch}}, \bibinfo {author} {\bibfnamefont
  {J.~I.}\ \bibnamefont {Cirac}},\ and\ \bibinfo {author} {\bibfnamefont
  {P.}~\bibnamefont {Zoller}},\ }\href@noop {} {\bibfield  {journal} {\bibinfo
  {journal} {Physical review letters}\ }\textbf {\bibinfo {volume} {87}},\
  \bibinfo {pages} {037901} (\bibinfo {year} {2001})}\BibitemShut {NoStop}%
\bibitem [{\citenamefont {Saffman}\ \emph {et~al.}(2010)\citenamefont
  {Saffman}, \citenamefont {Walker},\ and\ \citenamefont
  {M{\o}lmer}}]{saffman2010quantum}%
  \BibitemOpen
  \bibfield  {author} {\bibinfo {author} {\bibfnamefont {M.}~\bibnamefont
  {Saffman}}, \bibinfo {author} {\bibfnamefont {T.~G.}\ \bibnamefont
  {Walker}},\ and\ \bibinfo {author} {\bibfnamefont {K.}~\bibnamefont
  {M{\o}lmer}},\ }\href@noop {} {\bibfield  {journal} {\bibinfo  {journal}
  {Reviews of Modern Physics}\ }\textbf {\bibinfo {volume} {82}},\ \bibinfo
  {pages} {2313} (\bibinfo {year} {2010})}\BibitemShut {NoStop}%
\bibitem [{\citenamefont {Saffman}(2016)}]{saffman2016quantum}%
  \BibitemOpen
  \bibfield  {author} {\bibinfo {author} {\bibfnamefont {M.}~\bibnamefont
  {Saffman}},\ }\href@noop {} {\bibfield  {journal} {\bibinfo  {journal}
  {Journal of Physics B: Atomic, Molecular and Optical Physics}\ }\textbf
  {\bibinfo {volume} {49}},\ \bibinfo {pages} {202001} (\bibinfo {year}
  {2016})}\BibitemShut {NoStop}%
\bibitem [{\citenamefont {Wilk}\ \emph {et~al.}(2010)\citenamefont {Wilk},
  \citenamefont {Ga{\"e}tan}, \citenamefont {Evellin}, \citenamefont {Wolters},
  \citenamefont {Miroshnychenko}, \citenamefont {Grangier},\ and\ \citenamefont
  {Browaeys}}]{wilk2010entanglement}%
  \BibitemOpen
  \bibfield  {author} {\bibinfo {author} {\bibfnamefont {T.}~\bibnamefont
  {Wilk}}, \bibinfo {author} {\bibfnamefont {A.}~\bibnamefont {Ga{\"e}tan}},
  \bibinfo {author} {\bibfnamefont {C.}~\bibnamefont {Evellin}}, \bibinfo
  {author} {\bibfnamefont {J.}~\bibnamefont {Wolters}}, \bibinfo {author}
  {\bibfnamefont {Y.}~\bibnamefont {Miroshnychenko}}, \bibinfo {author}
  {\bibfnamefont {P.}~\bibnamefont {Grangier}},\ and\ \bibinfo {author}
  {\bibfnamefont {A.}~\bibnamefont {Browaeys}},\ }\href@noop {} {\bibfield
  {journal} {\bibinfo  {journal} {Phys. Rev. Lett.}\ }\textbf {\bibinfo
  {volume} {104}},\ \bibinfo {pages} {010502} (\bibinfo {year}
  {2010})}\BibitemShut {NoStop}%
\bibitem [{\citenamefont {Isenhower}\ \emph {et~al.}(2010)\citenamefont
  {Isenhower}, \citenamefont {Urban}, \citenamefont {Zhang}, \citenamefont
  {Gill}, \citenamefont {Henage}, \citenamefont {Johnson}, \citenamefont
  {Walker},\ and\ \citenamefont {Saffman}}]{isenhower2010demonstration}%
  \BibitemOpen
  \bibfield  {author} {\bibinfo {author} {\bibfnamefont {L.}~\bibnamefont
  {Isenhower}}, \bibinfo {author} {\bibfnamefont {E.}~\bibnamefont {Urban}},
  \bibinfo {author} {\bibfnamefont {X.-L.}\ \bibnamefont {Zhang}}, \bibinfo
  {author} {\bibfnamefont {A.}~\bibnamefont {Gill}}, \bibinfo {author}
  {\bibfnamefont {T.}~\bibnamefont {Henage}}, \bibinfo {author} {\bibfnamefont
  {T.-A.}\ \bibnamefont {Johnson}}, \bibinfo {author} {\bibfnamefont {T.-G.}\
  \bibnamefont {Walker}},\ and\ \bibinfo {author} {\bibfnamefont
  {M.}~\bibnamefont {Saffman}},\ }\href@noop {} {\bibfield  {journal} {\bibinfo
   {journal} {Phys. Rev. Lett.}\ }\textbf {\bibinfo {volume} {104}},\ \bibinfo
  {pages} {010503} (\bibinfo {year} {2010})}\BibitemShut {NoStop}%
\bibitem [{\citenamefont {Levine}\ \emph {et~al.}(2018)\citenamefont {Levine},
  \citenamefont {Keesling}, \citenamefont {Omran}, \citenamefont {Bernien},
  \citenamefont {Schwartz}, \citenamefont {Zibrov}, \citenamefont {Endres},
  \citenamefont {Greiner}, \citenamefont {Vuleti{\'c}},\ and\ \citenamefont
  {Lukin}}]{levine2018high}%
  \BibitemOpen
  \bibfield  {author} {\bibinfo {author} {\bibfnamefont {H.}~\bibnamefont
  {Levine}}, \bibinfo {author} {\bibfnamefont {A.}~\bibnamefont {Keesling}},
  \bibinfo {author} {\bibfnamefont {A.}~\bibnamefont {Omran}}, \bibinfo
  {author} {\bibfnamefont {H.}~\bibnamefont {Bernien}}, \bibinfo {author}
  {\bibfnamefont {S.}~\bibnamefont {Schwartz}}, \bibinfo {author}
  {\bibfnamefont {A.~S.}\ \bibnamefont {Zibrov}}, \bibinfo {author}
  {\bibfnamefont {M.}~\bibnamefont {Endres}}, \bibinfo {author} {\bibfnamefont
  {M.}~\bibnamefont {Greiner}}, \bibinfo {author} {\bibfnamefont
  {V.}~\bibnamefont {Vuleti{\'c}}},\ and\ \bibinfo {author} {\bibfnamefont
  {M.~D.}\ \bibnamefont {Lukin}},\ }\href@noop {} {\bibfield  {journal}
  {\bibinfo  {journal} {Phys. Rev. Lett.}\ }\textbf {\bibinfo {volume} {121}},\
  \bibinfo {pages} {123603} (\bibinfo {year} {2018})}\BibitemShut {NoStop}%
\bibitem [{\citenamefont {Bernien}\ \emph {et~al.}(2017)\citenamefont
  {Bernien}, \citenamefont {Schwartz}, \citenamefont {Keesling}, \citenamefont
  {Levine}, \citenamefont {Omran}, \citenamefont {Pichler}, \citenamefont
  {Choi}, \citenamefont {Zibrov}, \citenamefont {Endres}, \citenamefont
  {Greiner} \emph {et~al.}}]{bernien2017probing}%
  \BibitemOpen
  \bibfield  {author} {\bibinfo {author} {\bibfnamefont {H.}~\bibnamefont
  {Bernien}}, \bibinfo {author} {\bibfnamefont {S.}~\bibnamefont {Schwartz}},
  \bibinfo {author} {\bibfnamefont {A.}~\bibnamefont {Keesling}}, \bibinfo
  {author} {\bibfnamefont {H.}~\bibnamefont {Levine}}, \bibinfo {author}
  {\bibfnamefont {A.}~\bibnamefont {Omran}}, \bibinfo {author} {\bibfnamefont
  {H.}~\bibnamefont {Pichler}}, \bibinfo {author} {\bibfnamefont
  {S.}~\bibnamefont {Choi}}, \bibinfo {author} {\bibfnamefont {A.~S.}\
  \bibnamefont {Zibrov}}, \bibinfo {author} {\bibfnamefont {M.}~\bibnamefont
  {Endres}}, \bibinfo {author} {\bibfnamefont {M.}~\bibnamefont {Greiner}},
  \emph {et~al.},\ }\href@noop {} {\bibfield  {journal} {\bibinfo  {journal}
  {Nature}\ }\textbf {\bibinfo {volume} {551}},\ \bibinfo {pages} {579}
  (\bibinfo {year} {2017})}\BibitemShut {NoStop}%
\bibitem [{\citenamefont {Keesling}\ \emph {et~al.}(2019)\citenamefont
  {Keesling}, \citenamefont {Omran}, \citenamefont {Levine}, \citenamefont
  {Bernien}, \citenamefont {Pichler}, \citenamefont {Choi}, \citenamefont
  {Samajdar}, \citenamefont {Schwartz}, \citenamefont {Silvi}, \citenamefont
  {Sachdev} \emph {et~al.}}]{keesling2019quantum}%
  \BibitemOpen
  \bibfield  {author} {\bibinfo {author} {\bibfnamefont {A.}~\bibnamefont
  {Keesling}}, \bibinfo {author} {\bibfnamefont {A.}~\bibnamefont {Omran}},
  \bibinfo {author} {\bibfnamefont {H.}~\bibnamefont {Levine}}, \bibinfo
  {author} {\bibfnamefont {H.}~\bibnamefont {Bernien}}, \bibinfo {author}
  {\bibfnamefont {H.}~\bibnamefont {Pichler}}, \bibinfo {author} {\bibfnamefont
  {S.}~\bibnamefont {Choi}}, \bibinfo {author} {\bibfnamefont {R.}~\bibnamefont
  {Samajdar}}, \bibinfo {author} {\bibfnamefont {S.}~\bibnamefont {Schwartz}},
  \bibinfo {author} {\bibfnamefont {P.}~\bibnamefont {Silvi}}, \bibinfo
  {author} {\bibfnamefont {S.}~\bibnamefont {Sachdev}}, \emph {et~al.},\
  }\href@noop {} {\bibfield  {journal} {\bibinfo  {journal} {Nature}\ }\textbf
  {\bibinfo {volume} {568}},\ \bibinfo {pages} {207} (\bibinfo {year}
  {2019})}\BibitemShut {NoStop}%
\bibitem [{\citenamefont {Omran}\ \emph {et~al.}(2019)\citenamefont {Omran},
  \citenamefont {Levine}, \citenamefont {Keesling}, \citenamefont {Semeghini},
  \citenamefont {Wang}, \citenamefont {Ebadi}, \citenamefont {Bernien},
  \citenamefont {Zibrov}, \citenamefont {Pichler}, \citenamefont {Choi} \emph
  {et~al.}}]{omran2019generation}%
  \BibitemOpen
  \bibfield  {author} {\bibinfo {author} {\bibfnamefont {A.}~\bibnamefont
  {Omran}}, \bibinfo {author} {\bibfnamefont {H.}~\bibnamefont {Levine}},
  \bibinfo {author} {\bibfnamefont {A.}~\bibnamefont {Keesling}}, \bibinfo
  {author} {\bibfnamefont {G.}~\bibnamefont {Semeghini}}, \bibinfo {author}
  {\bibfnamefont {T.~T.}\ \bibnamefont {Wang}}, \bibinfo {author}
  {\bibfnamefont {S.}~\bibnamefont {Ebadi}}, \bibinfo {author} {\bibfnamefont
  {H.}~\bibnamefont {Bernien}}, \bibinfo {author} {\bibfnamefont {A.~S.}\
  \bibnamefont {Zibrov}}, \bibinfo {author} {\bibfnamefont {H.}~\bibnamefont
  {Pichler}}, \bibinfo {author} {\bibfnamefont {S.}~\bibnamefont {Choi}}, \emph
  {et~al.},\ }\href@noop {} {\bibfield  {journal} {\bibinfo  {journal}
  {Science}\ }\textbf {\bibinfo {volume} {365}},\ \bibinfo {pages} {570}
  (\bibinfo {year} {2019})}\BibitemShut {NoStop}%
\bibitem [{\citenamefont {Levine}\ \emph {et~al.}(2019)\citenamefont {Levine},
  \citenamefont {Keesling}, \citenamefont {Semeghini}, \citenamefont {Omran},
  \citenamefont {Wang}, \citenamefont {Ebadi}, \citenamefont {Bernien},
  \citenamefont {Greiner}, \citenamefont {Vuleti{\'c}}, \citenamefont {Pichler}
  \emph {et~al.}}]{levine2019parallel}%
  \BibitemOpen
  \bibfield  {author} {\bibinfo {author} {\bibfnamefont {H.}~\bibnamefont
  {Levine}}, \bibinfo {author} {\bibfnamefont {A.}~\bibnamefont {Keesling}},
  \bibinfo {author} {\bibfnamefont {G.}~\bibnamefont {Semeghini}}, \bibinfo
  {author} {\bibfnamefont {A.}~\bibnamefont {Omran}}, \bibinfo {author}
  {\bibfnamefont {T.~T.}\ \bibnamefont {Wang}}, \bibinfo {author}
  {\bibfnamefont {S.}~\bibnamefont {Ebadi}}, \bibinfo {author} {\bibfnamefont
  {H.}~\bibnamefont {Bernien}}, \bibinfo {author} {\bibfnamefont
  {M.}~\bibnamefont {Greiner}}, \bibinfo {author} {\bibfnamefont
  {V.}~\bibnamefont {Vuleti{\'c}}}, \bibinfo {author} {\bibfnamefont
  {H.}~\bibnamefont {Pichler}}, \emph {et~al.},\ }\href@noop {} {\bibfield
  {journal} {\bibinfo  {journal} {Phys. Rev. Let}\ }\textbf {\bibinfo {volume}
  {123}},\ \bibinfo {pages} {170503} (\bibinfo {year} {2019})}\BibitemShut
  {NoStop}%
\bibitem [{\citenamefont {Graham}\ \emph {et~al.}(2019)\citenamefont {Graham},
  \citenamefont {Kwon}, \citenamefont {Grinkemeyer}, \citenamefont {Marra},
  \citenamefont {Jiang}, \citenamefont {Lichtman}, \citenamefont {Sun},
  \citenamefont {Ebert},\ and\ \citenamefont {Saffman}}]{graham2019rydberg}%
  \BibitemOpen
  \bibfield  {author} {\bibinfo {author} {\bibfnamefont {T.}~\bibnamefont
  {Graham}}, \bibinfo {author} {\bibfnamefont {M.}~\bibnamefont {Kwon}},
  \bibinfo {author} {\bibfnamefont {B.}~\bibnamefont {Grinkemeyer}}, \bibinfo
  {author} {\bibfnamefont {Z.}~\bibnamefont {Marra}}, \bibinfo {author}
  {\bibfnamefont {X.}~\bibnamefont {Jiang}}, \bibinfo {author} {\bibfnamefont
  {M.}~\bibnamefont {Lichtman}}, \bibinfo {author} {\bibfnamefont
  {Y.}~\bibnamefont {Sun}}, \bibinfo {author} {\bibfnamefont {M.}~\bibnamefont
  {Ebert}},\ and\ \bibinfo {author} {\bibfnamefont {M.}~\bibnamefont
  {Saffman}},\ }\href@noop {} {\bibfield  {journal} {\bibinfo  {journal} {arXiv
  preprint arXiv:1908.06103}\ } (\bibinfo {year} {2019})}\BibitemShut {NoStop}%
\bibitem [{\citenamefont {Kr{\'a}l}\ \emph {et~al.}(2007)\citenamefont
  {Kr{\'a}l}, \citenamefont {Thanopulos},\ and\ \citenamefont
  {Shapiro}}]{kral2007colloquium}%
  \BibitemOpen
  \bibfield  {author} {\bibinfo {author} {\bibfnamefont {P.}~\bibnamefont
  {Kr{\'a}l}}, \bibinfo {author} {\bibfnamefont {I.}~\bibnamefont
  {Thanopulos}},\ and\ \bibinfo {author} {\bibfnamefont {M.}~\bibnamefont
  {Shapiro}},\ }\href@noop {} {\bibfield  {journal} {\bibinfo  {journal}
  {Reviews of modern physics}\ }\textbf {\bibinfo {volume} {79}},\ \bibinfo
  {pages} {53} (\bibinfo {year} {2007})}\BibitemShut {NoStop}%
\bibitem [{\citenamefont {Pohl}\ \emph {et~al.}(2003)\citenamefont {Pohl},
  \citenamefont {Pattard},\ and\ \citenamefont {Rost}}]{pohl2003plasma}%
  \BibitemOpen
  \bibfield  {author} {\bibinfo {author} {\bibfnamefont {T.}~\bibnamefont
  {Pohl}}, \bibinfo {author} {\bibfnamefont {T.}~\bibnamefont {Pattard}},\ and\
  \bibinfo {author} {\bibfnamefont {J.~M.}\ \bibnamefont {Rost}},\ }\href@noop
  {} {\bibfield  {journal} {\bibinfo  {journal} {Phys. Rev. A}\ }\textbf
  {\bibinfo {volume} {68}},\ \bibinfo {pages} {010703(R)} (\bibinfo {year}
  {2003})}\BibitemShut {NoStop}%
\bibitem [{\citenamefont {Johnson}\ and\ \citenamefont
  {Rolston}(2010)}]{johnson2010interactions}%
  \BibitemOpen
  \bibfield  {author} {\bibinfo {author} {\bibfnamefont {J.-E.}\ \bibnamefont
  {Johnson}}\ and\ \bibinfo {author} {\bibfnamefont {S.-L.}\ \bibnamefont
  {Rolston}},\ }\href@noop {} {\bibfield  {journal} {\bibinfo  {journal} {Phys.
  Rev. A}\ }\textbf {\bibinfo {volume} {82}},\ \bibinfo {pages} {033412}
  (\bibinfo {year} {2010})}\BibitemShut {NoStop}%
\bibitem [{\citenamefont {Zeiher}\ \emph {et~al.}(2016)\citenamefont {Zeiher},
  \citenamefont {Van~Bijnen}, \citenamefont {Schau{\ss}}, \citenamefont {Hild},
  \citenamefont {Choi}, \citenamefont {Pohl}, \citenamefont {Bloch},\ and\
  \citenamefont {Gross}}]{zeiher2016many}%
  \BibitemOpen
  \bibfield  {author} {\bibinfo {author} {\bibfnamefont {J.}~\bibnamefont
  {Zeiher}}, \bibinfo {author} {\bibfnamefont {R.}~\bibnamefont {Van~Bijnen}},
  \bibinfo {author} {\bibfnamefont {P.}~\bibnamefont {Schau{\ss}}}, \bibinfo
  {author} {\bibfnamefont {S.}~\bibnamefont {Hild}}, \bibinfo {author}
  {\bibfnamefont {J.-y.}\ \bibnamefont {Choi}}, \bibinfo {author}
  {\bibfnamefont {T.}~\bibnamefont {Pohl}}, \bibinfo {author} {\bibfnamefont
  {I.}~\bibnamefont {Bloch}},\ and\ \bibinfo {author} {\bibfnamefont
  {C.}~\bibnamefont {Gross}},\ }\href@noop {} {\bibfield  {journal} {\bibinfo
  {journal} {Nature Physics}\ }\textbf {\bibinfo {volume} {12}},\ \bibinfo
  {pages} {1095} (\bibinfo {year} {2016})}\BibitemShut {NoStop}%
\bibitem [{\citenamefont {Gil}\ \emph {et~al.}(2014)\citenamefont {Gil},
  \citenamefont {Mukherjee}, \citenamefont {Bridge}, \citenamefont {Jones},\
  and\ \citenamefont {Pohl}}]{gil2014spin}%
  \BibitemOpen
  \bibfield  {author} {\bibinfo {author} {\bibfnamefont {L.-I.-R.}\
  \bibnamefont {Gil}}, \bibinfo {author} {\bibfnamefont {R.}~\bibnamefont
  {Mukherjee}}, \bibinfo {author} {\bibfnamefont {E.-M.}\ \bibnamefont
  {Bridge}}, \bibinfo {author} {\bibfnamefont {M.-P.-A.}\ \bibnamefont
  {Jones}},\ and\ \bibinfo {author} {\bibfnamefont {T.}~\bibnamefont {Pohl}},\
  }\href@noop {} {\bibfield  {journal} {\bibinfo  {journal} {Phys. Rev. Lett.}\
  }\textbf {\bibinfo {volume} {112}},\ \bibinfo {pages} {103601} (\bibinfo
  {year} {2014})}\BibitemShut {NoStop}%
\bibitem [{\citenamefont {Norcia}\ \emph {et~al.}(2019)\citenamefont {Norcia},
  \citenamefont {Young}, \citenamefont {Eckner}, \citenamefont {Oelker},
  \citenamefont {Ye},\ and\ \citenamefont {Kaufman}}]{norcia2019seconds}%
  \BibitemOpen
  \bibfield  {author} {\bibinfo {author} {\bibfnamefont {M.~A.}\ \bibnamefont
  {Norcia}}, \bibinfo {author} {\bibfnamefont {A.~W.}\ \bibnamefont {Young}},
  \bibinfo {author} {\bibfnamefont {W.~J.}\ \bibnamefont {Eckner}}, \bibinfo
  {author} {\bibfnamefont {E.}~\bibnamefont {Oelker}}, \bibinfo {author}
  {\bibfnamefont {J.}~\bibnamefont {Ye}},\ and\ \bibinfo {author}
  {\bibfnamefont {A.~M.}\ \bibnamefont {Kaufman}},\ }\href@noop {} {\bibfield
  {journal} {\bibinfo  {journal} {Science}\ }\textbf {\bibinfo {volume}
  {366}},\ \bibinfo {pages} {93} (\bibinfo {year} {2019})}\BibitemShut
  {NoStop}%
\bibitem [{\citenamefont {Pupillo}\ \emph {et~al.}(2010)\citenamefont
  {Pupillo}, \citenamefont {Micheli}, \citenamefont {Boninsegni}, \citenamefont
  {Lesanovsky},\ and\ \citenamefont {Zoller}}]{pupillo2010strongly}%
  \BibitemOpen
  \bibfield  {author} {\bibinfo {author} {\bibfnamefont {G.}~\bibnamefont
  {Pupillo}}, \bibinfo {author} {\bibfnamefont {A.}~\bibnamefont {Micheli}},
  \bibinfo {author} {\bibfnamefont {M.}~\bibnamefont {Boninsegni}}, \bibinfo
  {author} {\bibfnamefont {I.}~\bibnamefont {Lesanovsky}},\ and\ \bibinfo
  {author} {\bibfnamefont {P.}~\bibnamefont {Zoller}},\ }\href@noop {}
  {\bibfield  {journal} {\bibinfo  {journal} {Phys. Rev. Lett.}\ }\textbf
  {\bibinfo {volume} {104}},\ \bibinfo {pages} {223002} (\bibinfo {year}
  {2010})}\BibitemShut {NoStop}%
\bibitem [{\citenamefont {Dauphin}\ \emph {et~al.}(2012)\citenamefont
  {Dauphin}, \citenamefont {Muller},\ and\ \citenamefont
  {Martin-Delgado}}]{dauphin2012rydberg}%
  \BibitemOpen
  \bibfield  {author} {\bibinfo {author} {\bibfnamefont {A.}~\bibnamefont
  {Dauphin}}, \bibinfo {author} {\bibfnamefont {M.}~\bibnamefont {Muller}},\
  and\ \bibinfo {author} {\bibfnamefont {M.-A.}\ \bibnamefont
  {Martin-Delgado}},\ }\href@noop {} {\bibfield  {journal} {\bibinfo  {journal}
  {Phys. Rev. A}\ }\textbf {\bibinfo {volume} {86}},\ \bibinfo {pages} {053618}
  (\bibinfo {year} {2012})}\BibitemShut {NoStop}%
\bibitem [{\citenamefont {Keating}\ \emph {et~al.}(2016)\citenamefont
  {Keating}, \citenamefont {Baldwin}, \citenamefont {Jau}, \citenamefont {Lee},
  \citenamefont {Biedermann},\ and\ \citenamefont
  {Deutsch}}]{keating2016arbitrary}%
  \BibitemOpen
  \bibfield  {author} {\bibinfo {author} {\bibfnamefont {T.}~\bibnamefont
  {Keating}}, \bibinfo {author} {\bibfnamefont {C.~H.}\ \bibnamefont
  {Baldwin}}, \bibinfo {author} {\bibfnamefont {Y.-Y.}\ \bibnamefont {Jau}},
  \bibinfo {author} {\bibfnamefont {J.}~\bibnamefont {Lee}}, \bibinfo {author}
  {\bibfnamefont {G.~W.}\ \bibnamefont {Biedermann}},\ and\ \bibinfo {author}
  {\bibfnamefont {I.~H.}\ \bibnamefont {Deutsch}},\ }\href@noop {} {\bibfield
  {journal} {\bibinfo  {journal} {Phys. Rev. Lett.}\ }\textbf {\bibinfo
  {volume} {117}},\ \bibinfo {pages} {213601} (\bibinfo {year}
  {2016})}\BibitemShut {NoStop}%
\bibitem [{\citenamefont {Jau}\ \emph {et~al.}(2016)\citenamefont {Jau},
  \citenamefont {Hankin}, \citenamefont {Keating}, \citenamefont {Deutsch},\
  and\ \citenamefont {Biedermann}}]{jau2016entangling}%
  \BibitemOpen
  \bibfield  {author} {\bibinfo {author} {\bibfnamefont {Y.-Y.}\ \bibnamefont
  {Jau}}, \bibinfo {author} {\bibfnamefont {A.}~\bibnamefont {Hankin}},
  \bibinfo {author} {\bibfnamefont {T.}~\bibnamefont {Keating}}, \bibinfo
  {author} {\bibfnamefont {I.}~\bibnamefont {Deutsch}},\ and\ \bibinfo {author}
  {\bibfnamefont {G.}~\bibnamefont {Biedermann}},\ }\href@noop {} {\bibfield
  {journal} {\bibinfo  {journal} {Nature Physics}\ }\textbf {\bibinfo {volume}
  {12}},\ \bibinfo {pages} {71} (\bibinfo {year} {2016})}\BibitemShut {NoStop}%
\bibitem [{\citenamefont {Lee}\ \emph {et~al.}(2017)\citenamefont {Lee},
  \citenamefont {Martin}, \citenamefont {Jau}, \citenamefont {Keating},
  \citenamefont {Deutsch},\ and\ \citenamefont
  {Biedermann}}]{lee2017demonstration}%
  \BibitemOpen
  \bibfield  {author} {\bibinfo {author} {\bibfnamefont {J.}~\bibnamefont
  {Lee}}, \bibinfo {author} {\bibfnamefont {M.~J.}\ \bibnamefont {Martin}},
  \bibinfo {author} {\bibfnamefont {Y.-Y.}\ \bibnamefont {Jau}}, \bibinfo
  {author} {\bibfnamefont {T.}~\bibnamefont {Keating}}, \bibinfo {author}
  {\bibfnamefont {I.~H.}\ \bibnamefont {Deutsch}},\ and\ \bibinfo {author}
  {\bibfnamefont {G.~W.}\ \bibnamefont {Biedermann}},\ }\href@noop {}
  {\bibfield  {journal} {\bibinfo  {journal} {Phys. Rev. A}\ }\textbf {\bibinfo
  {volume} {95}},\ \bibinfo {pages} {041801(R)} (\bibinfo {year}
  {2017})}\BibitemShut {NoStop}%
\bibitem [{\citenamefont {Keating}\ \emph {et~al.}(2015)\citenamefont
  {Keating}, \citenamefont {Cook}, \citenamefont {Hankin}, \citenamefont {Jau},
  \citenamefont {Biedermann},\ and\ \citenamefont
  {Deutsch}}]{keating2015robust}%
  \BibitemOpen
  \bibfield  {author} {\bibinfo {author} {\bibfnamefont {T.}~\bibnamefont
  {Keating}}, \bibinfo {author} {\bibfnamefont {R.~L.}\ \bibnamefont {Cook}},
  \bibinfo {author} {\bibfnamefont {A.~M.}\ \bibnamefont {Hankin}}, \bibinfo
  {author} {\bibfnamefont {Y.-Y.}\ \bibnamefont {Jau}}, \bibinfo {author}
  {\bibfnamefont {G.~W.}\ \bibnamefont {Biedermann}},\ and\ \bibinfo {author}
  {\bibfnamefont {I.~H.}\ \bibnamefont {Deutsch}},\ }\href@noop {} {\bibfield
  {journal} {\bibinfo  {journal} {Phys. Rev. A}\ }\textbf {\bibinfo {volume}
  {91}},\ \bibinfo {pages} {012337} (\bibinfo {year} {2015})}\BibitemShut
  {NoStop}%
\bibitem [{\citenamefont {M{\o}ller}\ \emph {et~al.}(2008)\citenamefont
  {M{\o}ller}, \citenamefont {Madsen},\ and\ \citenamefont
  {M{\o}lmer}}]{moller2008quantum}%
  \BibitemOpen
  \bibfield  {author} {\bibinfo {author} {\bibfnamefont {D.}~\bibnamefont
  {M{\o}ller}}, \bibinfo {author} {\bibfnamefont {L.~B.}\ \bibnamefont
  {Madsen}},\ and\ \bibinfo {author} {\bibfnamefont {K.}~\bibnamefont
  {M{\o}lmer}},\ }\href@noop {} {\bibfield  {journal} {\bibinfo  {journal}
  {Phys. Rev. Lett.}\ }\textbf {\bibinfo {volume} {100}},\ \bibinfo {pages}
  {170504} (\bibinfo {year} {2008})}\BibitemShut {NoStop}%
\bibitem [{\citenamefont {Rao}\ and\ \citenamefont
  {M{\o}lmer}(2014)}]{rao2014robust}%
  \BibitemOpen
  \bibfield  {author} {\bibinfo {author} {\bibfnamefont {D.~B.}\ \bibnamefont
  {Rao}}\ and\ \bibinfo {author} {\bibfnamefont {K.}~\bibnamefont
  {M{\o}lmer}},\ }\href@noop {} {\bibfield  {journal} {\bibinfo  {journal}
  {Phys. Rev. A}\ }\textbf {\bibinfo {volume} {89}},\ \bibinfo {pages}
  {030301(R)} (\bibinfo {year} {2014})}\BibitemShut {NoStop}%
\bibitem [{\citenamefont {Beterov}\ \emph {et~al.}(2016)\citenamefont
  {Beterov}, \citenamefont {Saffman}, \citenamefont {Yakshina}, \citenamefont
  {Tretyakov}, \citenamefont {Entin}, \citenamefont {Bergamini}, \citenamefont
  {Kuznetsova},\ and\ \citenamefont {Ryabtsev}}]{beterov2016two}%
  \BibitemOpen
  \bibfield  {author} {\bibinfo {author} {\bibfnamefont {I.}~\bibnamefont
  {Beterov}}, \bibinfo {author} {\bibfnamefont {M.}~\bibnamefont {Saffman}},
  \bibinfo {author} {\bibfnamefont {E.}~\bibnamefont {Yakshina}}, \bibinfo
  {author} {\bibfnamefont {D.}~\bibnamefont {Tretyakov}}, \bibinfo {author}
  {\bibfnamefont {V.}~\bibnamefont {Entin}}, \bibinfo {author} {\bibfnamefont
  {S.}~\bibnamefont {Bergamini}}, \bibinfo {author} {\bibfnamefont
  {E.}~\bibnamefont {Kuznetsova}},\ and\ \bibinfo {author} {\bibfnamefont
  {I.}~\bibnamefont {Ryabtsev}},\ }\href@noop {} {\bibfield  {journal}
  {\bibinfo  {journal} {Phys. Rev. A}\ }\textbf {\bibinfo {volume} {94}},\
  \bibinfo {pages} {062307} (\bibinfo {year} {2016})}\BibitemShut {NoStop}%
\bibitem [{\citenamefont {Wu}\ \emph {et~al.}(2017)\citenamefont {Wu},
  \citenamefont {Huang}, \citenamefont {Hu}, \citenamefont {Yang},\ and\
  \citenamefont {Zheng}}]{wu2017rydberg}%
  \BibitemOpen
  \bibfield  {author} {\bibinfo {author} {\bibfnamefont {H.}~\bibnamefont
  {Wu}}, \bibinfo {author} {\bibfnamefont {X.-R.}\ \bibnamefont {Huang}},
  \bibinfo {author} {\bibfnamefont {C.-S.}\ \bibnamefont {Hu}}, \bibinfo
  {author} {\bibfnamefont {Z.-B.}\ \bibnamefont {Yang}},\ and\ \bibinfo
  {author} {\bibfnamefont {S.-B.}\ \bibnamefont {Zheng}},\ }\href@noop {}
  {\bibfield  {journal} {\bibinfo  {journal} {Phys. Rev. A}\ }\textbf {\bibinfo
  {volume} {96}},\ \bibinfo {pages} {022321} (\bibinfo {year}
  {2017})}\BibitemShut {NoStop}%
\bibitem [{\citenamefont {Beterov}\ \emph {et~al.}(2018)\citenamefont
  {Beterov}, \citenamefont {Hamzina}, \citenamefont {Yakshina}, \citenamefont
  {Tretyakov}, \citenamefont {Entin},\ and\ \citenamefont
  {Ryabtsev}}]{beterov2018adiabatic}%
  \BibitemOpen
  \bibfield  {author} {\bibinfo {author} {\bibfnamefont {I.}~\bibnamefont
  {Beterov}}, \bibinfo {author} {\bibfnamefont {G.}~\bibnamefont {Hamzina}},
  \bibinfo {author} {\bibfnamefont {E.}~\bibnamefont {Yakshina}}, \bibinfo
  {author} {\bibfnamefont {D.}~\bibnamefont {Tretyakov}}, \bibinfo {author}
  {\bibfnamefont {V.}~\bibnamefont {Entin}},\ and\ \bibinfo {author}
  {\bibfnamefont {I.}~\bibnamefont {Ryabtsev}},\ }\href@noop {} {\bibfield
  {journal} {\bibinfo  {journal} {Phys. Rev. A}\ }\textbf {\bibinfo {volume}
  {97}},\ \bibinfo {pages} {032701} (\bibinfo {year} {2018})}\BibitemShut
  {NoStop}%
\bibitem [{\citenamefont {Saffman}\ \emph {et~al.}(2019)\citenamefont
  {Saffman}, \citenamefont {Beterov}, \citenamefont {Dalal}, \citenamefont
  {Paez},\ and\ \citenamefont {Sanders}}]{saffman2019symmetric}%
  \BibitemOpen
  \bibfield  {author} {\bibinfo {author} {\bibfnamefont {M.}~\bibnamefont
  {Saffman}}, \bibinfo {author} {\bibfnamefont {I.}~\bibnamefont {Beterov}},
  \bibinfo {author} {\bibfnamefont {A.}~\bibnamefont {Dalal}}, \bibinfo
  {author} {\bibfnamefont {E.}~\bibnamefont {Paez}},\ and\ \bibinfo {author}
  {\bibfnamefont {B.}~\bibnamefont {Sanders}},\ }\href@noop {} {\bibfield
  {journal} {\bibinfo  {journal} {arXiv preprint arXiv:1912.02977}\ } (\bibinfo
  {year} {2019})}\BibitemShut {NoStop}%
\bibitem [{\citenamefont {M{\o}lmer}\ and\ \citenamefont
  {S{\o}rensen}(1999)}]{molmer1999multiparticle}%
  \BibitemOpen
  \bibfield  {author} {\bibinfo {author} {\bibfnamefont {K.}~\bibnamefont
  {M{\o}lmer}}\ and\ \bibinfo {author} {\bibfnamefont {A.}~\bibnamefont
  {S{\o}rensen}},\ }\href@noop {} {\bibfield  {journal} {\bibinfo  {journal}
  {Phys. Rev. Lett.}\ }\textbf {\bibinfo {volume} {82}},\ \bibinfo {pages}
  {1835} (\bibinfo {year} {1999})}\BibitemShut {NoStop}%
\bibitem [{\citenamefont {S{\o}rensen}\ and\ \citenamefont
  {M{\o}lmer}(1999)}]{sorensen1999quantum}%
  \BibitemOpen
  \bibfield  {author} {\bibinfo {author} {\bibfnamefont {A.}~\bibnamefont
  {S{\o}rensen}}\ and\ \bibinfo {author} {\bibfnamefont {K.}~\bibnamefont
  {M{\o}lmer}},\ }\href@noop {} {\bibfield  {journal} {\bibinfo  {journal}
  {Phys. Rev. Lett.}\ }\textbf {\bibinfo {volume} {82}},\ \bibinfo {pages}
  {1971} (\bibinfo {year} {1999})}\BibitemShut {NoStop}%
\bibitem [{\citenamefont {Jaksch}\ \emph {et~al.}(2000)\citenamefont {Jaksch},
  \citenamefont {Cirac}, \citenamefont {Zoller}, \citenamefont {Rolston},
  \citenamefont {C{\^o}t{\'e}},\ and\ \citenamefont {Lukin}}]{jaksch2000fast}%
  \BibitemOpen
  \bibfield  {author} {\bibinfo {author} {\bibfnamefont {D.}~\bibnamefont
  {Jaksch}}, \bibinfo {author} {\bibfnamefont {J.}~\bibnamefont {Cirac}},
  \bibinfo {author} {\bibfnamefont {P.}~\bibnamefont {Zoller}}, \bibinfo
  {author} {\bibfnamefont {S.}~\bibnamefont {Rolston}}, \bibinfo {author}
  {\bibfnamefont {R.}~\bibnamefont {C{\^o}t{\'e}}},\ and\ \bibinfo {author}
  {\bibfnamefont {M.}~\bibnamefont {Lukin}},\ }\href@noop {} {\bibfield
  {journal} {\bibinfo  {journal} {Phys. Rev. Lett.}\ }\textbf {\bibinfo
  {volume} {85}},\ \bibinfo {pages} {2208} (\bibinfo {year}
  {2000})}\BibitemShut {NoStop}%
\bibitem [{\citenamefont {Covey}\ \emph {et~al.}(2019)\citenamefont {Covey},
  \citenamefont {Madjarov}, \citenamefont {Cooper},\ and\ \citenamefont
  {Endres}}]{covey20192000}%
  \BibitemOpen
  \bibfield  {author} {\bibinfo {author} {\bibfnamefont {J.~P.}\ \bibnamefont
  {Covey}}, \bibinfo {author} {\bibfnamefont {I.~S.}\ \bibnamefont {Madjarov}},
  \bibinfo {author} {\bibfnamefont {A.}~\bibnamefont {Cooper}},\ and\ \bibinfo
  {author} {\bibfnamefont {M.}~\bibnamefont {Endres}},\ }\href@noop {}
  {\bibfield  {journal} {\bibinfo  {journal} {Phys. Rev. Lett.}\ }\textbf
  {\bibinfo {volume} {122}},\ \bibinfo {pages} {173201} (\bibinfo {year}
  {2019})}\BibitemShut {NoStop}%
\bibitem [{\citenamefont {Saskin}\ \emph {et~al.}(2019)\citenamefont {Saskin},
  \citenamefont {Wilson}, \citenamefont {Grinkemeyer},\ and\ \citenamefont
  {Thompson}}]{saskin2019narrow}%
  \BibitemOpen
  \bibfield  {author} {\bibinfo {author} {\bibfnamefont {S.}~\bibnamefont
  {Saskin}}, \bibinfo {author} {\bibfnamefont {J.-T.}\ \bibnamefont {Wilson}},
  \bibinfo {author} {\bibfnamefont {B.}~\bibnamefont {Grinkemeyer}},\ and\
  \bibinfo {author} {\bibfnamefont {J.~D.}\ \bibnamefont {Thompson}},\
  }\href@noop {} {\bibfield  {journal} {\bibinfo  {journal} {Phys. Rev. Lett.}\
  }\textbf {\bibinfo {volume} {122}},\ \bibinfo {pages} {143002} (\bibinfo
  {year} {2019})}\BibitemShut {NoStop}%
\bibitem [{\citenamefont {Zhang}\ \emph {et~al.}(2003)\citenamefont {Zhang},
  \citenamefont {Vala}, \citenamefont {Sastry},\ and\ \citenamefont
  {Whaley}}]{zhang2003geometric}%
  \BibitemOpen
  \bibfield  {author} {\bibinfo {author} {\bibfnamefont {J.}~\bibnamefont
  {Zhang}}, \bibinfo {author} {\bibfnamefont {J.}~\bibnamefont {Vala}},
  \bibinfo {author} {\bibfnamefont {S.}~\bibnamefont {Sastry}},\ and\ \bibinfo
  {author} {\bibfnamefont {K.~B.}\ \bibnamefont {Whaley}},\ }\href@noop {}
  {\bibfield  {journal} {\bibinfo  {journal} {Phys. Rev. A}\ }\textbf {\bibinfo
  {volume} {67}},\ \bibinfo {pages} {042313} (\bibinfo {year}
  {2003})}\BibitemShut {NoStop}%
\bibitem [{\citenamefont {Martin}\ \emph {et~al.}(2018)\citenamefont {Martin},
  \citenamefont {Biedermann}, \citenamefont {Jau}, \citenamefont {Lee},\ and\
  \citenamefont {Deutsch}}]{martin2018cphase}%
  \BibitemOpen
  \bibfield  {author} {\bibinfo {author} {\bibfnamefont {M.~J.}\ \bibnamefont
  {Martin}}, \bibinfo {author} {\bibfnamefont {G.}~\bibnamefont {Biedermann}},
  \bibinfo {author} {\bibfnamefont {Y.-Y.}\ \bibnamefont {Jau}}, \bibinfo
  {author} {\bibfnamefont {J.}~\bibnamefont {Lee}},\ and\ \bibinfo {author}
  {\bibfnamefont {I.}~\bibnamefont {Deutsch}},\ }\href@noop {} {\emph {\bibinfo
  {title} {A CPHASE gate between Rydberg-dressed neutral atoms.}}},\ \bibinfo
  {type} {Tech. Rep.}\ (\bibinfo  {institution} {Sandia National Lab.(SNL-NM),
  Albuquerque, NM (United States)},\ \bibinfo {year} {2018})\BibitemShut
  {NoStop}%
\bibitem [{\citenamefont {Pedersen}\ \emph {et~al.}(2007)\citenamefont
  {Pedersen}, \citenamefont {M{\o}ller},\ and\ \citenamefont
  {M{\o}lmer}}]{pedersen2007fidelity}%
  \BibitemOpen
  \bibfield  {author} {\bibinfo {author} {\bibfnamefont {L.~H.}\ \bibnamefont
  {Pedersen}}, \bibinfo {author} {\bibfnamefont {N.~M.}\ \bibnamefont
  {M{\o}ller}},\ and\ \bibinfo {author} {\bibfnamefont {K.}~\bibnamefont
  {M{\o}lmer}},\ }\href@noop {} {\bibfield  {journal} {\bibinfo  {journal}
  {Physics Letters A}\ }\textbf {\bibinfo {volume} {367}},\ \bibinfo {pages}
  {47} (\bibinfo {year} {2007})}\BibitemShut {NoStop}%
\bibitem [{\citenamefont {Zhang}\ \emph {et~al.}(2012)\citenamefont {Zhang},
  \citenamefont {Gill}, \citenamefont {Isenhower}, \citenamefont {Walker},\
  and\ \citenamefont {Saffman}}]{zhang2012fidelity}%
  \BibitemOpen
  \bibfield  {author} {\bibinfo {author} {\bibfnamefont {X.}~\bibnamefont
  {Zhang}}, \bibinfo {author} {\bibfnamefont {A.}~\bibnamefont {Gill}},
  \bibinfo {author} {\bibfnamefont {L.}~\bibnamefont {Isenhower}}, \bibinfo
  {author} {\bibfnamefont {T.}~\bibnamefont {Walker}},\ and\ \bibinfo {author}
  {\bibfnamefont {M.}~\bibnamefont {Saffman}},\ }\href@noop {} {\bibfield
  {journal} {\bibinfo  {journal} {Phys. Rev. A}\ }\textbf {\bibinfo {volume}
  {85}},\ \bibinfo {pages} {042310} (\bibinfo {year} {2012})}\BibitemShut
  {NoStop}%
\bibitem [{\citenamefont {Walker}\ and\ \citenamefont
  {Saffman}(2012)}]{walker2012entanglement}%
  \BibitemOpen
  \bibfield  {author} {\bibinfo {author} {\bibfnamefont {T.~G.}\ \bibnamefont
  {Walker}}\ and\ \bibinfo {author} {\bibfnamefont {M.}~\bibnamefont
  {Saffman}},\ }in\ \href@noop {} {\emph {\bibinfo {booktitle} {Advances in
  Atomic, Molecular, and Optical Physics}}},\ Vol.~\bibinfo {volume} {61}\
  (\bibinfo  {publisher} {Elsevier},\ \bibinfo {year} {2012})\ pp.\ \bibinfo
  {pages} {81--115}\BibitemShut {NoStop}%
\bibitem [{\citenamefont {de~L{\'e}s{\'e}leuc}\ \emph
  {et~al.}(2018{\natexlab{b}})\citenamefont {de~L{\'e}s{\'e}leuc},
  \citenamefont {Barredo}, \citenamefont {Lienhard}, \citenamefont {Browaeys},\
  and\ \citenamefont {Lahaye}}]{de2018analysis}%
  \BibitemOpen
  \bibfield  {author} {\bibinfo {author} {\bibfnamefont {S.}~\bibnamefont
  {de~L{\'e}s{\'e}leuc}}, \bibinfo {author} {\bibfnamefont {D.}~\bibnamefont
  {Barredo}}, \bibinfo {author} {\bibfnamefont {V.}~\bibnamefont {Lienhard}},
  \bibinfo {author} {\bibfnamefont {A.}~\bibnamefont {Browaeys}},\ and\
  \bibinfo {author} {\bibfnamefont {T.}~\bibnamefont {Lahaye}},\ }\href@noop {}
  {\bibfield  {journal} {\bibinfo  {journal} {Phys. Rev. A}\ }\textbf {\bibinfo
  {volume} {97}},\ \bibinfo {pages} {053803} (\bibinfo {year}
  {2018}{\natexlab{b}})}\BibitemShut {NoStop}%
\bibitem [{\citenamefont {Beterov}\ \emph {et~al.}(2009)\citenamefont
  {Beterov}, \citenamefont {Ryabtsev}, \citenamefont {Tretyakov},\ and\
  \citenamefont {Entin}}]{beterov2009quasiclassical}%
  \BibitemOpen
  \bibfield  {author} {\bibinfo {author} {\bibfnamefont {I.-I.}\ \bibnamefont
  {Beterov}}, \bibinfo {author} {\bibfnamefont {I.-I.}\ \bibnamefont
  {Ryabtsev}}, \bibinfo {author} {\bibfnamefont {D.-B.}\ \bibnamefont
  {Tretyakov}},\ and\ \bibinfo {author} {\bibfnamefont {V.-M.}\ \bibnamefont
  {Entin}},\ }\href@noop {} {\bibfield  {journal} {\bibinfo  {journal} {Phys.
  Rev. A}\ }\textbf {\bibinfo {volume} {79}},\ \bibinfo {pages} {052504}
  (\bibinfo {year} {2009})}\BibitemShut {NoStop}%
\bibitem [{\citenamefont {Goerz}\ \emph {et~al.}(2014)\citenamefont {Goerz},
  \citenamefont {Halperin}, \citenamefont {Aytac}, \citenamefont {Koch},\ and\
  \citenamefont {Whaley}}]{goerz2014robustness}%
  \BibitemOpen
  \bibfield  {author} {\bibinfo {author} {\bibfnamefont {M.~H.}\ \bibnamefont
  {Goerz}}, \bibinfo {author} {\bibfnamefont {E.~J.}\ \bibnamefont {Halperin}},
  \bibinfo {author} {\bibfnamefont {J.~M.}\ \bibnamefont {Aytac}}, \bibinfo
  {author} {\bibfnamefont {C.~P.}\ \bibnamefont {Koch}},\ and\ \bibinfo
  {author} {\bibfnamefont {K.~B.}\ \bibnamefont {Whaley}},\ }\href@noop {}
  {\bibfield  {journal} {\bibinfo  {journal} {Phys. Rev. A}\ }\textbf {\bibinfo
  {volume} {90}},\ \bibinfo {pages} {032329} (\bibinfo {year}
  {2014})}\BibitemShut {NoStop}%
\end{thebibliography}%
\bibliographystyle{apsrev4-2}

\end{document}